\documentclass[runningheads,a4paper]{llncs}
\usepackage[T1]{fontenc}
\usepackage{graphicx}
\usepackage{enumitem}
\usepackage{amsmath}
\usepackage{amssymb} 
\usepackage{rotating}
\usepackage{pdflscape}

\usepackage{subcaption}

\usepackage{booktabs}
\usepackage{tabularx}
\usepackage{ragged2e}
\newcolumntype{Y}{>{\RaggedRight\arraybackslash}X}

\begin{document}
\title{Semantic Non-Fungibility and Violations of the Law of One Price in Prediction Markets}
%
\titlerunning{Liquidity Fragmentation in Prediction Markets}
\author{Jonas Gebele \and
Florian Matthes
}
\authorrunning{Gebele and Matthes}
\institute{Technical University of Munich, Munich, Germany\\
\email{\{jonas.gebele, matthes\}@tum.de}}
\maketitle
\begin{abstract}
Prediction markets are designed to aggregate dispersed information about future events, yet today’s ecosystem is fragmented across heterogeneous operator-run platforms and blockchain-based protocols that independently list economically identical events. In the absence of a shared notion of event identity, liquidity fails to pool across venues, arbitrage becomes capital-intensive or unenforceable, and prices systematically violate the Law of One Price. As a result, market prices reflect platform-local beliefs rather than a single, globally aggregated probability, undermining the core information-aggregation function of prediction markets.
We address this gap by introducing a semantic alignment framework that makes cross-platform event identity explicit through joint analysis of natural-language descriptions, resolution semantics, and temporal scope. Applying this framework, we construct the first human-validated, cross-platform dataset of aligned prediction markets, covering over 100,000 events across ten major venues from 2018 to 2025.
Using this dataset, we show that roughly 6\% of all events are concurrently listed across platforms and that semantically equivalent markets exhibit persistent execution-aware price deviations of 2–4\% on average, even in highly liquid and information-rich settings. These mispricings give rise to persistent cross-platform arbitrage opportunities driven by structural frictions rather than informational disagreement. Overall, our results demonstrate that semantic non-fungibility is a fundamental barrier to price convergence, and that resolving event identity is a prerequisite for prediction markets to aggregate information at a global scale.

\end{abstract}

\section{Introduction}

Prediction markets allow participants to trade contingent claims whose prices reflect collective beliefs about uncertain future events. Their ability to aggregate dispersed information has been documented in a wide range of settings, from the Iowa Electronic Markets (IEM)~\cite{IEM2025} to internal corporate forecasting systems at Google~\cite{Cowgill2009}, Hewlett--Packard~\cite{Plott2002}, Intel~\cite{Gillen2013}, and Eli Lilly~\cite{Polgreen2007}. Across these deployments, market prices have been shown to incorporate private information and to generate forecasts that are competitive with, and often superior to, alternative prediction mechanisms.

\noindent
Yet as prediction markets have expanded across a heterogeneous ecosystem of operator-run platforms and decentralized blockchain protocols, a fundamental structural limitation has emerged: the absence of a shared, machine-verifiable notion of \emph{event identity}. In contrast to traditional financial assets, which are defined by globally unique identifiers (e.g., ISINs, futures contract codes, or ERC-20 token addresses), prediction markets specify contingent claims through platform-specific natural-language descriptions together with resolution rules, oracle sources, and cutoff times. As a consequence, determining whether two markets promise the same payoff is non-trivial and cannot be automated reliably.

This lack of event identity renders economically equivalent contingent claims \emph{semantically non-fungible}: they cannot be treated as interchangeable assets, netted across venues, or arbitraged without committing capital until resolution. For example, Kalshi’s market \emph{“Which party will win the Senate?”}\footnote{\url{https://kalshi.com/markets/controls/senate-winner/controls-2024}} and Polymarket’s market \emph{“Senate control after the 2024 election?”}\footnote{\url{https://polymarket.com/event/which-party-will-control-the-us-senate-after-the-2024-election}} refer to the same underlying proposition, yet no existing mechanism can verify their payoff equivalence in a way that allows positions to be offset across platforms.

The economic consequences of missing semantic interoperability are structural rather than incidental. In the absence of a common event identity,
\begin{enumerate}[leftmargin=2.0em, itemsep=3pt]
\item liquidity fragments across platforms instead of pooling,
\item prices reflect platform-local beliefs rather than global information, and
\item economically equivalent claims trade at persistent, non-aligned prices.
\end{enumerate}
\noindent
Crucially, these divergences do not reflect disagreement about the underlying event. Instead, they arise from a structural failure of enforceability: arbitrage positions cannot be netted across venues and must be held until resolution, hindering arbitrage from equalizing prices and leading to systematic violations of the Law of One Price even in liquid, information-rich settings.

\medskip
\noindent
Despite its importance, this problem has received little attention in the academic literature. Existing work on prediction markets focuses almost exclusively on \emph{single-platform} settings, examining price discovery~\cite{Ng2025}, wash trading~\cite{Sirolly2025}, or within-platform arbitrage~\cite{saguillo_et_al:LIPIcs.AFT.2025.27}. By construction, these studies treat markets referring to the same real-world event as already aligned and therefore abstract away from cross-platform event identity. In the absence of a framework for establishing event identity across venues, it is not possible to quantify liquidity fragmentation, characterize cross-platform information aggregation, or evaluate arbitrage opportunities that span multiple markets.

\begin{figure*}[t!]
    \centering
    \includegraphics[width=\textwidth]{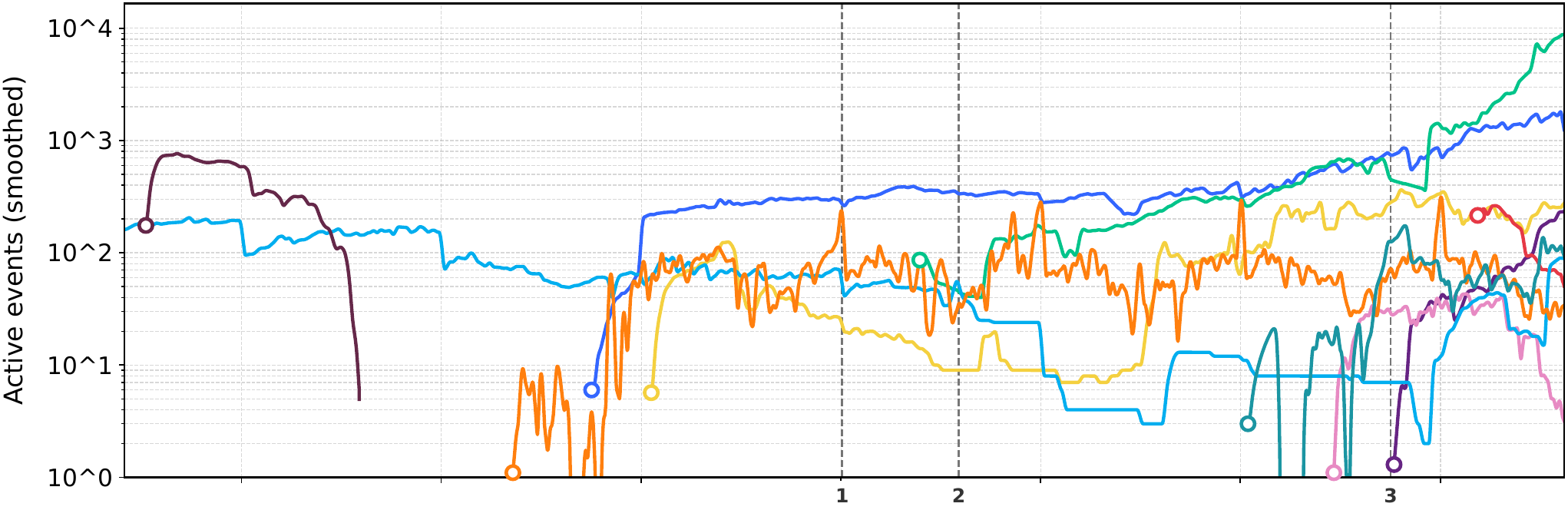}
    \vspace{0.5em}
    \includegraphics[width=\textwidth]{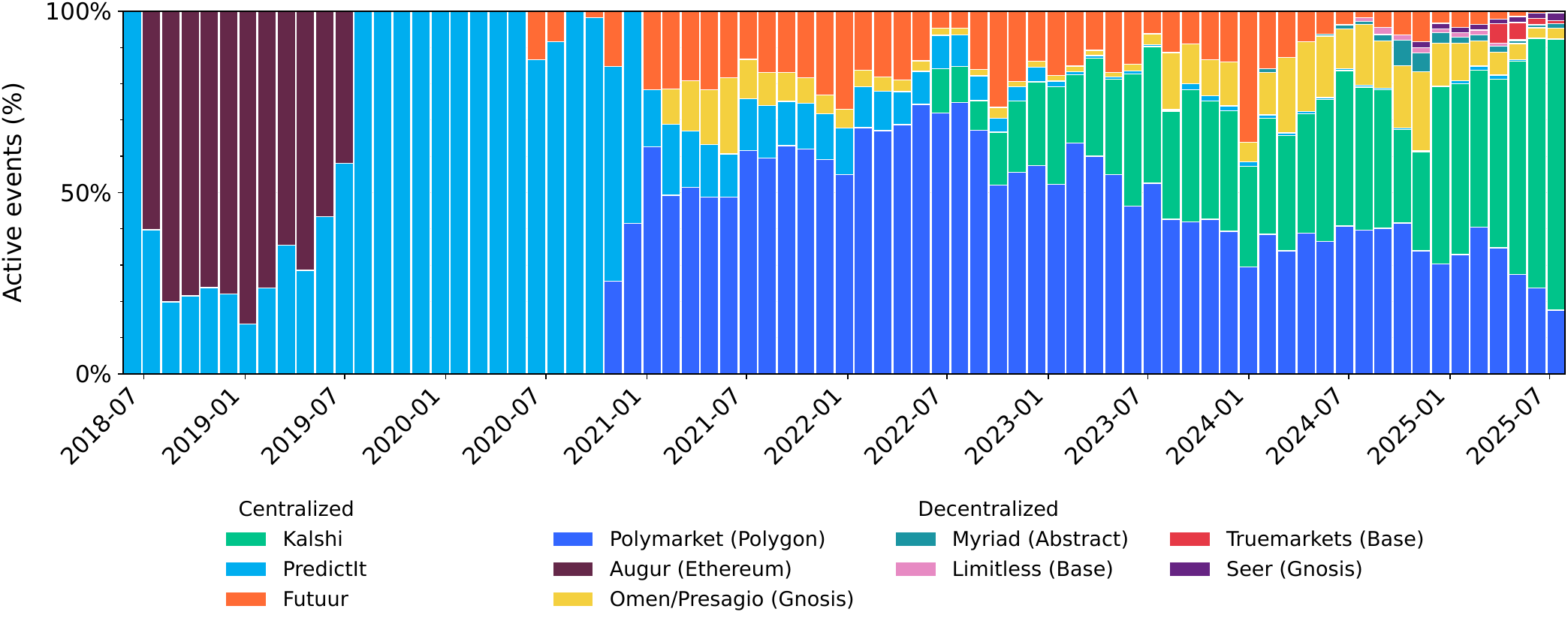}
    \caption{
        \textbf{Top:} Number of active prediction-market events over time (log scale; 5-day moving average). Circles indicate the launch dates of individual platforms. Vertical dashed lines \textbf{(1)}--\textbf{(3)} mark major U.S.\ regulatory interventions affecting market availability: \textbf{(1)} the CFTC’s enforcement action against Polymarket in January~2022; \textbf{(2)} the withdrawal of PredictIt’s no-action letter in August~2022; and \textbf{(3)} the 2024 federal court rulings allowing Kalshi’s Congressional Control markets to proceed during appeal.
        \textbf{Bottom:} Platform-level composition of active events over time, shown as relative shares (5-week resampled).
    }
    \label{fig:open-markets-combined}
\end{figure*}

\smallskip
\noindent
The scale of this challenge has grown rapidly. Figure~\ref{fig:open-markets-combined} illustrates the expansion of the prediction-market ecosystem since 2018 across a heterogeneous mix of operator-run platforms~\cite{kalshi,PredictIt,Futuur} and decentralized blockchain-based protocols~\cite{augurproject_github,polymarket,omen_eth,Limitless,MyriadMarkets,Seer.pm,TrueMarkets}. More recently, large consumer financial platforms such as Robinhood, DraftKings, and Coinbase have begun offering prediction-market products~\cite{Conlon2025,Cunningham2025,Saini2025}, further broadening participation and increasing market coverage. While blockchain-based markets have enabled open participation, stablecoin settlement, and programmable liquidity, they have also introduced greater heterogeneity in event specification, oracle design, and market microstructure. As a result, the number of independently listed markets has increased substantially, expanding the scope for semantic duplication and cross-platform fragmentation.

\noindent
This paper addresses the absence of semantic foundations needed to reason consistently across prediction-market platforms. We present the first ecosystem-scale empirical analysis of prediction markets spanning both operator-run and blockchain-based platforms, explicitly modeling event identity and its economic consequences. Our contributions are threefold:

\begin{enumerate}[leftmargin=2.2em, itemsep=3pt, topsep=2pt]
    \item \textbf{Cross-Platform Event Identification at Scale.}
    We develop a scalable framework for identifying semantically equivalent and subset-related markets across heterogeneous platforms. Applied to over 100{,}000 events, the framework makes event identity explicit and enables systematic measurement of semantic fragmentation at the ecosystem level.
    \item \textbf{A Human-Validated Cross-Platform Event Dataset.}
    We construct and release the first human-validated dataset aligning prediction-market events across platforms, including market descriptions, resolution semantics, oracle sources, and verified equivalence relations. The dataset provides a foundation for empirical and theoretical work on event identity and market interoperability.
    \item \textbf{Quantifying Fragmentation and Cross-Platform Price Divergence.}
    Using the aligned dataset, we quantify the magnitude and persistence of price divergence across semantically equivalent markets. We document systematic deviations from execution-adjusted parity and show that cross-platform arbitrage opportunities arise from structural limits to enforceability rather than informational disagreement.
\end{enumerate}

\section{Background}

Prediction markets allow participants to trade fully collateralized contingent claims on future events. While their core economic structure is shared across platforms, prediction-market systems differ substantially in how events are specified, resolved, and priced. These differences give rise to \emph{semantic fragmentation}: the absence of a machine-verifiable notion of event identity across platforms.

This section first establishes a common economic abstraction for prediction-market claims, and then summarizes the platform-level design choices that generate semantic fragmentation and motivate our cross-platform analysis.

\subsection{Binary Contingent Claims}

Prediction markets most commonly implement contingent claims as binary outcome contracts. Let $X \in \{0,1\}$ denote the realized outcome of the underlying event. A YES-share pays $\pi = X$ at resolution, while a NO-share pays $1-X$. Let $p_Y$ and $p_N$ denote the market prices of the YES- and NO-shares, respectively. Holding both claims until resolution therefore yields a deterministic unit payoff, implying the parity condition
\[
p_Y + p_N = 1.
\]

Categorical markets decompose naturally into collections of mutually exclusive binary claims, enabling economic comparability even when platforms adopt different outcome granularities (e.g., a single multi-outcome election market versus multiple binary markets). Under standard assumptions, the price of a YES-share has a probabilistic interpretation,
\[
p_Y \approx \mathbb{E}[X],
\]
which underlies essentially all empirical analyses of prediction markets.

\subsection{Event Representation}

Prediction-market platforms share no common identifier or formal schema for real-world events. Instead, each market specifies its underlying condition through a platform-specific natural-language description, supplemented by metadata such as resolution rules, oracle references, and cutoff times. These descriptions determine the event’s semantics but are expressed informally and lack a shared, machine-verifiable representation across platforms. Operator-run platforms typically expose this information through proprietary APIs, while decentralized platforms embed it directly into smart contracts at deployment.

\medskip
\noindent
Distinct from event specification is the \emph{representation of claims}. Operator-run platforms record trader positions in internal ledgers, whereas decentralized platforms generally tokenize outcome claims on-chain. A widely used design is the \emph{Gnosis Conditional Token Framework (CTF)}~\cite{Gnosis2019}, which represents each condition as ERC-1155 tokens whose identifier is a hash of the oracle address, a platform-defined \texttt{questionId}, and the number of outcomes. While this construction enables composability within a platform, it also implies that any differences in event specification generate distinct, non-fungible identifiers.

\medskip
\noindent
Across platforms, market instantiation differs along three main dimensions:
\begin{enumerate}[itemsep=2pt, topsep=2pt]
    \item \textbf{Oracle source} (e.g., reporter-based, committee-based, optimistic oracles),
    \item \textbf{Collateral and payout asset} (e.g., USD, USDC, DAI),
    \item \textbf{Outcome granularity} (binary, categorical, or scalar).
\end{enumerate}

\smallskip
\noindent
These representation choices determine how a market is instantiated and traded, but its economic meaning at settlement ultimately depends on platform-specific resolution rules. As a result, differences in representation interact with heterogeneous resolution semantics, introducing an additional layer of semantic heterogeneity across platforms.

\subsection{Resolution Semantics}
\label{subsec:resolution_semantic}

Even markets with nearly identical wording may resolve differently because their truth conditions are governed by platform-specific resolution semantics. These semantics vary along three primary dimensions:
\begin{enumerate}[itemsep=2pt, topsep=2pt]
    \item \textbf{Reference source:} the external authority used to determine the outcome,
    \item \textbf{Temporal scope:} the precise cutoff time at which the outcome is evaluated,
    \item \textbf{Exception rules:} invalidation criteria, dispute and appeal procedures, multi-stage finality, and amendment mechanisms.
\end{enumerate}

A concrete example illustrates the economic relevance of these differences.   
Kalshi’s market \emph{“Highest temperature in NYC?”}\footnote{\url{https://kalshi.com/markets/kxhighny/highest-temperature-in-nyc}} resolves using data from NOAA’s Central Park weather station, whereas the corresponding market on Polymarket\footnote{\url{https://polymarket.com/event/highest-temperature-in-nyc-on-december-2}} references measurements from LaGuardia Airport. These locations frequently record different temperatures. As a result, markets that appear nearly equivalent at the textual level encode distinct resolution semantics and correspond to meaningfully different contingent claims.

\subsection{Market Microstructure}

Once outcomes are defined, platforms differ in how trader beliefs are translated into observable market prices. Two broad families of pricing mechanisms dominate prediction markets.

\paragraph{Automated market makers (AMMs).}
AMMs determine prices through a deterministic pricing rule, typically derived from a cost function or a reserve invariant. Prices adjust mechanically in response to trades, with market depth and price sensitivity governed by platform-specific liquidity parameters. Two canonical AMM designs commonly used in prediction markets are the Logarithmic Market Scoring Rule (LMSR), which offers bounded loss to the market maker~\cite{repec:buc:jpredm:v:1:y:2007:i:1:p:3-15}, and constant-product market makers (CPMMs), in which prices are determined by the ratio of outcome-share reserves~\cite{Adams2018}. While these mechanisms differ in design, both induce deterministic but platform-specific price dynamics.

\paragraph{Continuous limit order books (CLOBs).}
Other platforms employ continuous limit order books, in which traders submit buy and sell orders at discrete prices and quantities. Trades occur when orders cross, producing bid--ask spreads, tick-size constraints, and depth-dependent execution prices. As a result, observed prices (e.g., mid-quotes) reflect order-book microstructure rather than a continuous implied probability.

\medskip
Differences in pricing mechanisms, liquidity provision, fee structures, tick sizes, and access constraints can sustain persistent price differences even when underlying beliefs are identical. These microstructure frictions interact with the semantic fragmentation described above, jointly shaping the cross-platform price disparities analyzed in the remainder of the paper. This motivates the semantic-alignment framework introduced in the sections that follow.

\section{Formal Model of Markets and Arbitrage}
\label{sec.formal_market_model}

Although prediction-market platforms vary widely in how markets are specified, traded, and settled, they all exchange collateral-backed binary contingent claims whose prices reflect beliefs about future states. We introduce a unified abstraction that isolates this shared economic structure across platforms and provides a common basis for cross-platform price comparison and arbitrage analysis.

\paragraph{Normalization and modeling assumptions.}
To facilitate comparison across heterogeneous markets, we adopt the following assumptions.  
(i) All prices are expressed in a common USD-equivalent unit.  
(ii) Arbitrage conditions are evaluated on synchronized price snapshots, such that offsetting trades are feasible within a bounded latency window (accounted for empirically in~\S\ref{sec:results}).  
(iii) The representative arbitrageur can legally and operationally hold all required positions, including both YES and NO claims (regulatory segmentation is discussed in~\S\ref{sec.case_study}).  
(iv) Arbitrage positions are held until market resolution.  
(v) Each market resolves according to its stated resolution rules; dispute and appeal processes are assumed to terminate.

\subsection{Platforms, Events, and Markets}

Let $\mathcal{P}=\{P_1,\dots,P_\ell\}$ denote the set of platforms. Each platform lists a collection of real-world events, where each event gives rise to one or more tradable markets. We restrict attention to \emph{binary} markets, which pay a unit payoff contingent on whether a specified condition is satisfied. Multi-outcome (categorical) markets are accommodated by representing each possible outcome as a separate binary claim that pays one if and only if that outcome occurs. Scalar markets (e.g., temperatures or inflation levels) are excluded, as they would require discretization schemes beyond our scope.

For any binary market $m$, let $p_Y(m)\in[0,1]$ denote the price of a YES-claim and let $p_N(m)\in[0,1]$ denote the price of the corresponding NO-claim.

\smallskip
\noindent
Execution frictions such as bid--ask spreads, platform fees, gas costs, tick-size constraints, and slippage are summarized by a one-sided, non-negative parameter
\[
\delta(m)\ge 0.
\]
We apply $\delta(m)$ uniformly to both YES and NO positions in all no-arbitrage bounds. In the formal model, it serves as a compact abstraction of execution frictions; in the empirical analysis, we apply platform-specific, conservative estimates (see Appendix~\ref{app:execution_costs_section}).

\subsection{Single-Platform Conditional Arbitrage}

A binary market pays one unit at resolution to either the YES or the NO position. In the absence of arbitrage, the combined acquisition cost of holding both outcome claims must therefore be close to one. Allowing for execution frictions, no-arbitrage requires
\[
1-\delta(m)\;\le\;p_Y(m)+p_N(m)\;\le\;1+\delta(m).
\]
If this condition is violated, a trader can secure a deterministic profit by simultaneously purchasing (or selling) both the YES and NO claims, locking in a fixed payoff at resolution regardless of the event outcome. This within-market parity condition serves as a baseline for the cross-platform arbitrage conditions introduced below.

\medskip
\noindent
In practice, such parity violations are rare, as most platforms employ a unified pricing mechanism for both outcome positions that enforces this constraint mechanically.

\subsection{Cross-Platform Conditional Arbitrage}

Consider two semantically equivalent binary markets $m_i$ and $m_j$ listed on platforms $P_i$ and $P_j$, respectively, which refer to the same underlying real-world proposition. Let
\[
\Delta_{ij} = \delta(m_i) + \delta(m_j)
\]
denote the total execution friction incurred when opening offsetting positions across both platforms.

Because exactly one of the YES or NO positions pays out at resolution, acquiring a YES position on one platform and the corresponding NO position on the other yields a unit payoff regardless of the event outcome. In the absence of arbitrage, the total acquisition cost of such a cross-platform bundle must therefore be close to one. Allowing for execution frictions, cross-platform no-arbitrage requires
\[
1-\Delta_{ij}
\;\le\;
\min\!\bigl\{ p_Y(m_i)+p_N(m_j),\; p_Y(m_j)+p_N(m_i) \bigr\}
\;\le\;
1+\Delta_{ij}.
\]

\noindent
If this condition is violated, a trader can secure a deterministic profit by purchasing the underpriced bundle and holding the position until resolution. This condition is the cross-platform analogue of within-market YES--NO parity.

\medskip
\noindent
More generally, markets may be related by logical implication rather than strict equivalence; we formalize this case below using subset relations.

\subsection{Single-Platform Negative-Risk Arbitrage}
\label{sec:negativerisk}

Negative-risk arbitrage arises when multiple binary markets traded on the same platform are linked by a single underlying event. Although these markets trade independently, their payoffs are mutually exclusive and collectively exhaustive. Let $M=\{m_1,\dots,m_n\}$ denote a set of binary markets corresponding to the possible outcomes of one event, such that exactly one market resolves YES at settlement (e.g., individual contracts on each team winning the World Cup).

Since exactly one YES-claim pays out, holding one unit of each YES-claim in $M$ yields a guaranteed unit payoff at resolution, while holding one unit of each NO-claim yields a guaranteed payoff of $n-1$. Allowing for execution frictions, the absence of arbitrage therefore requires
\[
\sum_{m\in M}\!\bigl(p_Y(m)+\delta(m)\bigr)\;\ge\;1,
\qquad
\sum_{m\in M}\!\bigl(p_N(m)+\delta(m)\bigr)\;\ge\;n-1.
\]

\noindent
A strict violation of either inequality creates a deterministic arbitrage opportunity: the corresponding bundle of YES-claims or NO-claims can be acquired below its guaranteed payoff and held until resolution.

\subsection{Cross-Platform Negative-Risk Arbitrage}

Platforms often encode the same multi-outcome event at different levels of granularity. While one venue may list each outcome as an individual binary market, another may aggregate multiple low-probability outcomes into a single composite outcome (e.g., an ``other'' contract covering all remaining candidates). As a result, no single platform may induce a complete partition of the event, even though collections of markets drawn from multiple platforms may jointly exhaust all possible outcomes. To reason about such cases, we introduce a shared \emph{atomic outcome space}.

\paragraph{Atomic outcome space.}
Let $\Omega=\{\omega_1,\dots,\omega_k\}$ denote the mutually exclusive and collectively exhaustive outcomes of a given event, serving as a common reference space across platforms. For each binary market $m$, define its YES-region $f(m)\subseteq\Omega$ as the set of outcomes under which $m$ resolves YES. We assume resolution is deterministic conditional on each platform’s stated rules, so that each market can be represented as an indicator over $\Omega$. In practice, $\Omega$ and the mappings $f(m)$ are approximated using the semantic-alignment pipeline described in~\S\ref{sec:pipeline}.

\paragraph{Cross-platform partitions.}
A collection of markets $M$ drawn from multiple platforms forms a \emph{valid cross-platform partition} of an event if the YES-regions of its constituent markets satisfy the following conditions over the atomic outcome space $\Omega$:
\[
\begin{aligned}
\textbf{(Completeness)} 
&\quad \bigcup_{m\in M} f(m)=\Omega,\\[3pt]
\textbf{(Mutual Exclusivity)} 
&\quad f(m)\cap f(m')=\emptyset \quad \text{for all } m\neq m',\\[3pt]
\textbf{(Unique Resolution)} 
&\quad \forall\,\omega\in\Omega:\;\exists!\,m\in M \text{ such that } \omega\in f(m).
\end{aligned}
\]
Under these conditions, exactly one market in $M$ resolves YES at settlement. Economically, the collection behaves like a single categorical market, despite being distributed across platforms.

\paragraph{Parity conditions.}
For a valid cross-platform partition $M$, the categorical no-arbitrage conditions extend directly across platforms. Holding one unit of each YES-claim in $M$ yields a guaranteed unit payoff at resolution, while holding one unit of each NO-claim yields a guaranteed payoff of $|M|-1$. Allowing for execution frictions, the absence of arbitrage therefore requires
\[
\sum_{m\in M}\!\bigl(p_Y(m)+\delta(m)\bigr)\;\ge\;1,
\qquad
\sum_{m\in M}\!\bigl(p_N(m)+\delta(m)\bigr)\;\ge\;|M|-1.
\]

\noindent
A strict violation of either inequality gives rise to deterministic cross-platform negative-risk arbitrage. For example, if the YES-bundle is underpriced, the arbitrage profit is
\[
\pi^{\mathrm{cross}}_{\mathrm{YES}}
=
1-\sum_{m\in M}\bigl(p_Y(m)+\delta(m)\bigr),
\]
and analogously for the NO-bundle. Thus, once markets are aligned semantically, negative-risk arbitrage generalizes seamlessly from single-platform to cross-platform settings.

\subsection{Subset Relations and Conditional Coverage}

Subset relations arise when the YES-region of one market is strictly contained in that of another. Formally, a market $m_s$ is a \emph{subset} of a \emph{superset} market $m_S$ if $f(m_s)\subset f(m_S)$. Such relations encode logical implication between markets and enable conditional arbitrage.

A YES position on the superset market and NO on the subset market spans the entire atomic outcome space and therefore yields a deterministic unit payoff at resolution.

For example, a market resolving YES if a Republican wins the 2028 U.S.\ presidential election is a strict superset of a market resolving YES if J.D.\ Vance wins the election. Holding YES on the former and NO on the latter guarantees a unit payoff regardless of the election outcome.


\medskip
\noindent
\textbf{Conceptual vs.\ empirical equivalence.}
The formal model assumes perfect knowledge of each market’s resolution function $r_m:\Omega\to\{Y,N\}$. Empirically, we approximate YES-regions $f(m)$ using LLM-based semantic analysis of textual descriptions, cutoff times, and resolution metadata (Section~\ref{sec:pipeline}). Residual approximation error arises from paraphrasing ambiguity and heterogeneous platform semantics and is quantified in~\S\ref{sec:results}.

\section{Data and Semantic-Matching Framework}

Cross-platform analysis of prediction markets requires first establishing when markets listed on different venues refer to the same underlying event. We therefore construct a unified cross-platform dataset and introduce a scalable semantic-matching framework that infers equivalence and subset relations between markets based on their descriptions, resolution semantics, and temporal scope. This framework makes event identity explicit and underpins all subsequent measurements.

\subsection{Dataset Construction}
We construct a multi-venue dataset of prediction markets spanning both centralized operator-run and decentralized platforms through \textbf{August~2025}. The unit of observation is an \emph{event together with its associated binary outcome set}, as defined in~\S\ref{sec.formal_market_model}. Platform-specific markets are treated as distinct instantiations of this underlying event.

\paragraph{Platform inclusion.}

We include a platform if it lists \emph{at least fifty markets} with individual lifetime trading volume exceeding \textbf{\$500}. This threshold restricts attention to economically meaningful venues, ensuring that 

\begin{enumerate}[label=(\roman*), leftmargin=1.8em]
\item observed prices reflect genuine monetary exposure rather than symbolic or reputation-based incentives, and
\item small experimental or short-lived deployments do not distort ecosystem-wide statistics.
\end{enumerate}

\noindent
We \emph{exclude}:
\begin{itemize}[leftmargin=1.6em]
\item \textbf{play-money or reputation-based systems} such as \emph{Metaculus}~\cite{2025d} and \emph{Manifold}~\cite{2025c}, whose prices do not reflect real economic risk;
\item \textbf{venues whose prices mechanically replicate external reference odds}, including sportsbooks (e.g., \emph{Azuro}~\cite{2025}, \emph{SX.bet}~\cite{2025a}, \emph{Betfair}~\cite{2025b}) or platforms offering synthetic option-style payoff replication.
In these systems, prices arise from replication rather than independent information aggregation.
\end{itemize}

\paragraph{Temporal scope.}
For each platform, we include all available historical market data through \textbf{August~15--22, 2025}, subject to platform-specific availability constraints and filtering procedures summarized in Table~\ref{tab:data-acquisition}.

\paragraph{Data acquisition and normalization.}

Data from decentralized platforms are obtained via public RPC endpoints and blockchain indexing services, while data from centralized operator-run platforms are collected through official APIs. For order-book venues, we use reported mid-quote prices; for AMM-based markets, prices are reconstructed directly from on-chain reserves at the transaction level. All timestamps are standardized to UTC and all prices are converted to USD-equivalent units to ensure cross-platform comparability. We restrict the analysis to non-sports markets and apply deterministic filters to exclude markets with missing temporal information, zero trading volume, or internally inconsistent records. Table~\ref{tab:data-acquisition} summarizes data sources, schemas, and cleaning procedures.

\paragraph{Dataset summary.}

The final dataset comprises over \textbf{100{,}000 unique events} drawn from \emph{Polymarket}, \emph{Omen/Presagio}, \emph{Augur~v1}, \emph{Limitless}, \emph{Myriad}, \emph{Seer}, \emph{Truemarkets}, \emph{Kalshi}, \emph{Futuur}, and \emph{PredictIt}. Cross-validation against publicly available platform analytics confirms consistency across data sources. The dataset will be made publicly available upon publication.

\subsection{Cross-Platform Event Identity}

Prediction markets specify contingent claims using heterogeneous natural-language descriptions combined with platform-specific oracle rules, cutoff times, and scope qualifiers. As a result, determining whether two markets refer to the same underlying proposition requires reasoning about event semantics rather than surface-level text similarity.

\noindent
Following~\S\ref{sec.formal_market_model}, we model each market $m$ as a resolution function
\[
r_m : \Omega \rightarrow \{Y,N\},
\]
where $\Omega$ denotes the atomic outcome space of the underlying event. We define the \emph{YES-region} of a market as $f(m) \subseteq \Omega$, the set of atomic states under which the market resolves affirmatively.

This representation induces three economically relevant relations between any two markets $m_1$ and $m_2$. Table~\ref{tab:relations} illustrates these relations relative to a fixed reference market.  Markets are \emph{equivalent} if they resolve identically in all states ($f(m_1)=f(m_2)$), \emph{subset-related} if one market’s YES-region is strictly contained in the other ($f(m_1)\subset f(m_2)$), and \emph{independent} otherwise.

\begin{table}[h]
    \centering
        \caption{Semantic relations between markets, illustrated relative to the reference market ``Donald Trump elected as next president in 2024''.}
        \label{tab:relations}
        \renewcommand{\arraystretch}{1.15}
        \begin{tabular}{lll}
        \toprule
        \textbf{Relation} & \textbf{Formal condition} & \textbf{Example market} \\
        \midrule
        Equivalent 
        & $f(m_1) = f(m_2)$ 
        & ``Donald Trump wins 2024 U.S.\ presidential election'' \\[4pt]
        Subset 
        & $f(m_1) \subset f(m_2)$ 
        & ``Trump receives at least 300 Electoral College votes'' \\[4pt]
        Independent 
        & otherwise 
        & ``Democrats win the 2024 U.S.\ Senate majority'' \\
        \bottomrule
    \end{tabular}
\end{table}

Subset relations arise when one proposition logically implies another, for example due to stricter thresholds, narrower scope, or earlier evaluation times. Identifying such relations requires semantic and contextual reasoning beyond text similarity, including careful interpretation of resolution rules and cutoff semantics.

\subsection{Semantic-Matching Pipeline}
\label{sec:pipeline}

Naively comparing all pairs among roughly 100{,}000 events yields on the order of $10^{10}$ candidate pairs and is computationally infeasible. We therefore employ a layered pipeline that progressively narrows the candidate set before applying detailed semantic verification.

\subsubsection{1. Structural Filtering}

The first stage eliminates pairs that cannot plausibly refer to the same underlying event. All filters are conservative and designed to preserve essentially all true matches:
\begin{itemize}[leftmargin=1.5em]
    \item \textbf{Cross-platform constraint:} only markets listed on different platforms are compared.
    \item \textbf{Category matching:} each market is assigned to one of twenty semantic categories using an LLM-based classifier (Appendix~\ref{app:categories}); only markets within the same category are considered.
    \item \textbf{Temporal overlap:} markets must have \emph{any non-empty overlap} in their validity windows.
\end{itemize}

\noindent
After structural filtering, the candidate set for each market is typically reduced to approximately $10^{4}$ pairs.

\subsubsection{2. Semantic Retrieval via Embedding Similarity}

We represent each market using a vector embedding constructed from its title, description, outcome labels, and available resolution metadata, using the OpenAI \texttt{text-embedding-3-large} model (3072 dimensions). Within each structurally compatible subset, we retrieve the $k=20$ nearest neighbors under cosine similarity.

The choice of $k=20$ is empirically justified. Appendix~\ref{app:rank_distribution} shows that more than 99.9\% of all verified equivalent and subset relations appear within the top twenty nearest neighbors, bounding recall loss from embedding-based retrieval.

\begin{figure}[ht]
    \centering
    \includegraphics[width=\linewidth]{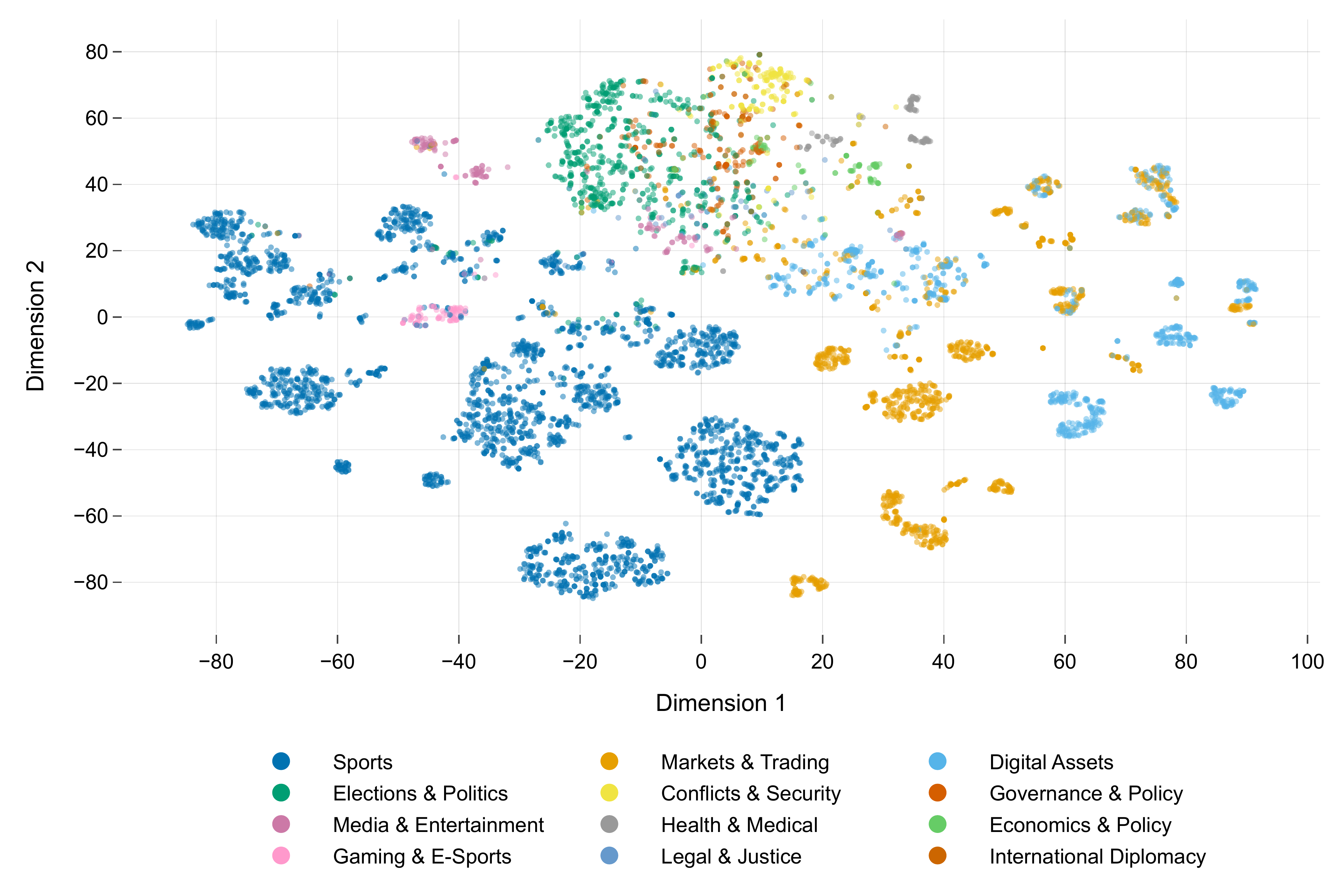}
    \caption{
        \textbf{t\textnormal{-}SNE projection of 7{,}000 sampled event embeddings.}
        Colors denote assigned semantic categories.  
        The emergence of coherent topical clusters indicates that embedding similarity captures meaningful event-level semantics, supporting its use as a high-recall candidate-retrieval signal.
    }
    \label{fig:tsne_embeddings}
\end{figure}

Figure~\ref{fig:tsne_embeddings} provides a qualitative visualization of the embedding space using a t-SNE projection of 7{,}000 sampled events. The emergence of coherent topical clusters indicates that embedding similarity captures meaningful event-level semantics and serves as a high-recall candidate-generation mechanism.

\begin{figure*}[htbp]
    \centering
    \begin{subfigure}[t]{0.49\textwidth}
        \centering
        \includegraphics[width=\linewidth]{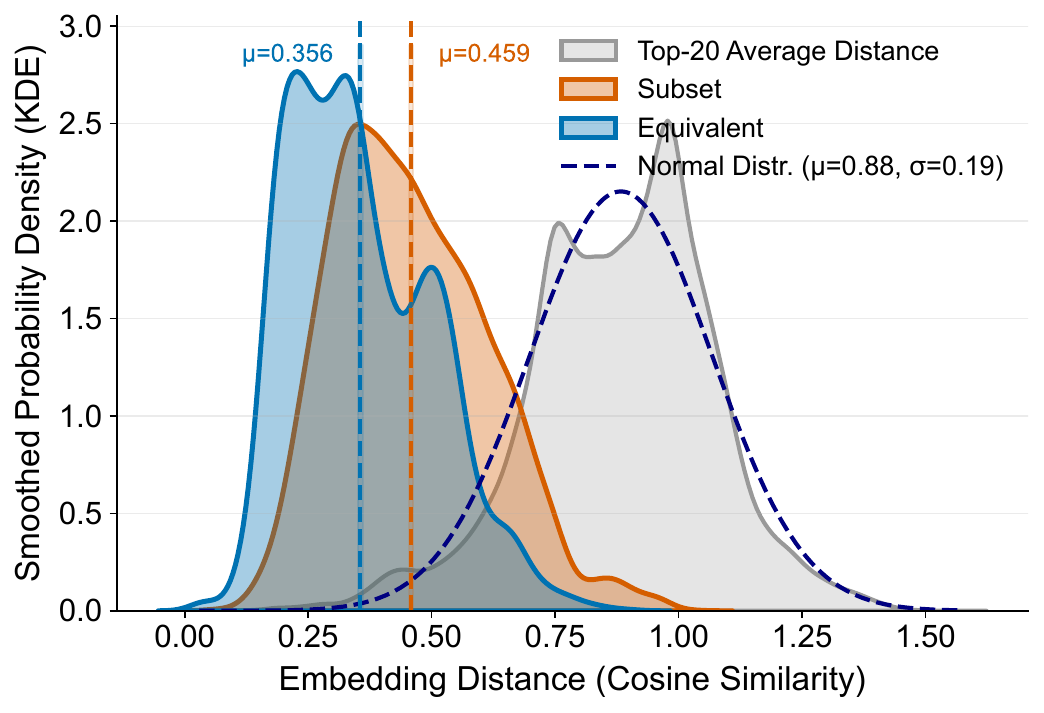}
        \caption{
                Distribution of embedding distances for validated equivalent, subset-related, and unrelated market pairs, relative to the top-$20$ retrieval baseline.
        }
        \label{fig:distance_distributions}
    \end{subfigure}
    \hfill
    \begin{subfigure}[t]{0.49\textwidth}
        \centering
        \includegraphics[width=\linewidth]{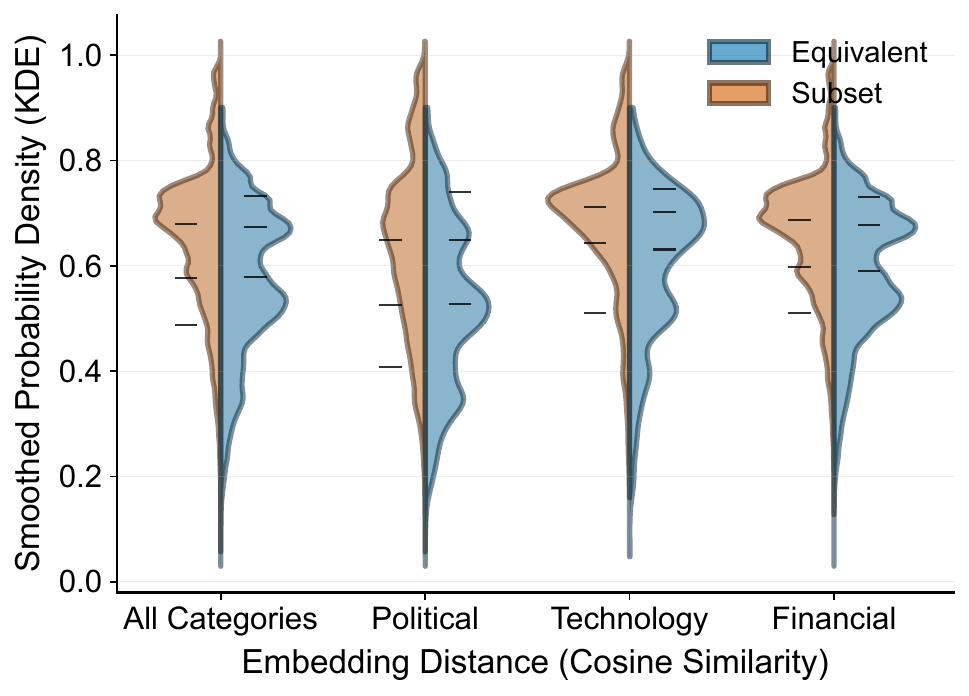}
        \caption{
                Split violin plots of embedding distances by relation type for selected domains, with internal quartile markers (25\%, 50\%, 75\%).
        }
        \label{fig:violin_split_subset}
    \end{subfigure}

    \vspace{-0.5em}
    \caption{
            Comparative distribution of embedding distances across semantic relation types and market domains.
    }
    \label{fig:embedding_distance_overview}
\end{figure*}

Figure~\ref{fig:embedding_distance_overview} quantifies this separation. Embedding distances are smallest for validated equivalent pairs, slightly larger for subset relations, and substantially larger for unrelated candidates retrieved within the top-$20$. This pattern holds consistently across domains and confirms that embedding similarity is informative but insufficient on its own, motivating the logical verification stage that follows.

\subsubsection{3. Logical Verification (Resolution-Semantics Reasoning)}

The final stage performs detailed semantic verification using all available market metadata, including cutoff times, oracle identities, dispute procedures, scope qualifiers, and conditional clauses. Verification is carried out in two LLM-based passes. First, a high-recall plausibility check filters out candidate pairs that are clearly semantically incompatible. Second, a structured LLM-based comparison explicitly compares the implied YES-regions $f(m)$ to determine equivalence or subset relations. Prompt templates for both stages are provided in Appendix~\ref{app:filtering_prompt} and Appendix \ref{app:equivalence_prompt} and \ref{app:subset_prompt}.

\subsubsection{Arbitrage Construction}

Given the identified semantic relations, we construct arbitrage \emph{positions} by combining binary outcome claims (YES/NO tokens or ledger positions) and evaluating their execution-adjusted parity conditions as defined in~\S\ref{sec.formal_market_model}.

\paragraph{Conditional-token arbitrage.}

For both equivalent and subset-related markets, the semantic relations identified by the verification stage directly determine a pair of offsetting YES/NO positions that spans the atomic outcome space~$\Omega$ (cf.~\S\ref{sec.formal_market_model}). Price divergence is therefore evaluated solely via the execution-adjusted parity conditions defined in the formal model.

\paragraph{Negative-risk arbitrage.}
For multi-outcome events, negative-risk arbitrage arises from cross-platform partitions of the atomic outcome space~$\Omega$ (see~\S\ref{sec.formal_market_model}). We identify such partitions via a constrained substitution procedure. Starting from a baseline multi-outcome event on one platform, we retrieve semantically equivalent binary outcome markets listed on other platforms and generate candidate partitions by substituting a small number of baseline outcomes with their cross-platform counterparts. Candidate partitions are retained only if they satisfy exact coverage and mutual exclusivity over~$\Omega$. Due to computational constraints, substitutions are limited to \textbf{2--4} outcomes per partition, and for each baseline event we evaluate at most the top \textbf{1{,}000} substitution permutations ranked by total trading volume.

\subsection{Evaluation Protocol and Error Analysis}

We evaluate the semantic-matching pipeline using stratified human annotation, stage-wise performance accounting, and a targeted error analysis focused on residual misclassifications.

\paragraph{Human annotation and agreement.}
We conduct human validation in two regimes.
First, to assess end-to-end labeling on the natural candidate distribution (dominated by unrelated pairs), one human annotator independently labeled a stratified sample of 1{,}000 candidate pairs. In this setting, agreement was effectively perfect.\footnote{This sample is heavily skewed toward true negatives; agreement is therefore primarily a check against spurious matches.}
Second, to assess the \emph{logical verification} stage where non-trivial semantic decisions arise, one human annotator independently labeled 1{,}000 verification-stage instances drawn from the post-retrieval candidate set. Inter-rater agreement in this regime was high ($\kappa = 0.94$), indicating consistent judgments on equivalence and subset relations across platforms and categories.

\paragraph{Stage-wise performance.}
Across the validated relations, the pipeline achieves high recall and low false-positive rates:
\begin{itemize}[leftmargin=1.5em]
    \item Structural filtering preserves $100\%$ of validated true relations.
    \item Embedding retrieval with $k=20$ captures $99.9\%$ of verified equivalent and subset relations.
    \item LLM-based logical verification reduces the false-positive rate to below $2\%$.
\end{itemize}

\paragraph{Residual errors and consistency checks.}
Residual errors are primarily false negatives arising from stale markets, highly specific resolution semantics, or implicit scope qualifiers. As a consistency check, markets labeled semantically equivalent were compared ex post using their stated resolution criteria and, where available, realized outcomes; we observed as the only exception the New York weather-station case discussed in Section~\ref{subsec:resolution_semantic}, which reflects genuinely distinct semantics and was not correctly classified by both the human annotator and the framework.

\paragraph{Scalability.}

After structural pruning, runtime scales approximately linearly in dataset size. Applying the full pipeline to over 100{,}000 events required on the order of \textbf{200 million tokens} across all LLM calls. While computationally intensive, this cost reflects one-time processing and demonstrates that ecosystem-scale semantic alignment is feasible with current models.

\section{Results}
\label{sec:results}

We first quantify the extent and structural properties of cross-platform semantic overlap in the prediction-market ecosystem. We then examine how this overlap is distributed across platforms and how it evolves over time. Finally, we measure the associated price divergence and characterize the persistence and magnitude of execution-aware arbitrage opportunities.

\subsection{Extent and Structure of Cross-Platform Semantic Overlap}

Out of 102{,}275 events, our semantic-matching pipeline identifies approximately \textbf{6\%} events that are involved in at least one cross-platform semantic relation. Although limited in number, these events represent nearly \textbf{10\%} of total event-days, reflecting systematic differences in market duration. Whereas the median prediction market lasts just over one day, semantically linked events typically remain active for several weeks. Cross-platform fragmentation therefore concentrates in long-lived, high-salience events that are concurrently listed across multiple venues. Most linked events appear on only one or two platforms, while a small heavy tail spans many venues, with a handful of events listed on up to eight platforms.

Decomposing this overlap reveals a highly structured arbitrage landscape. We identify 1{,}501 equivalence classes comprising 6{,}709 relations, 1{,}645 subset-related event sets comprising 6{,}421 relations, and 1{,}123 negative-risk constructions spanning 2{,}771 markets. The resulting relation topology closely mirrors logical semantics: equivalence classes form fully connected components in which all markets referencing the same proposition are mutually linked, whereas subset relations exhibit asymmetric hub-and-spoke structures connecting a single superset market to multiple stricter formulations.

Negative-risk constructions never occur in isolation and are almost always embedded within equivalence or subset structures. These differences are reflected in internal connectivity: equivalence classes are the densest, averaging 4.5 relations per class, followed by subset structures at 3.9 relations, while negative-risk constructions remain tightly constrained by logical partitioning, spanning 2.5 markets on average.

\subsection{Platform-Level Structure of Semantic Overlap}

\begin{figure}[t]
  \centering
  \includegraphics[width=0.75\textwidth]{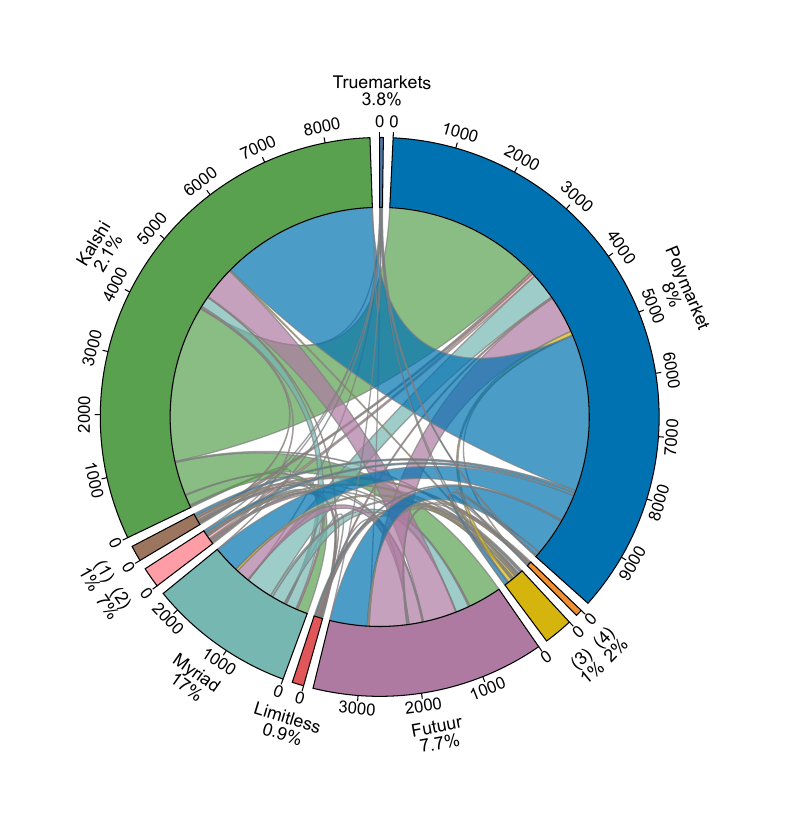}
  \caption{%
    Chord diagram of equivalent relations between prediction-market platforms.
    Arc size is proportional to the total number of equivalence relations involving each platform.
    Percentages indicate the share of a platform’s events that have an equivalent counterpart elsewhere.
    Ribbon thickness represents the number of matched market pairs, and ribbon color encodes direction, indicating which platform listed the event first.
    Short labels: (1) PredictIt, (2) Seer, (3) Omen, (4) Augur.
  }
  \label{fig:chord-diagram}
\end{figure}

Figure~\ref{fig:chord-diagram} summarizes the platform-level topology and directionality of equivalent-market relations.

\textbf{Kalshi} and \textbf{Polymarket} dominate the network, accounting for the majority of equivalence relations. Despite similar prominence, their exposure differs substantially: approximately 8\% of Polymarket’s markets have an equivalent counterpart on another platform, compared to about 2\% for Kalshi. This difference reflects listing scope rather than semantic isolation. Kalshi’s broad event coverage dilutes the relative share of duplicated markets, whereas Polymarket’s more selective listings concentrate on high-salience events that are frequently co-listed elsewhere. Directionality between the two platforms is nearly balanced, consistent with competitive co-listing rather than one-sided propagation.

Other platforms occupy more asymmetric positions that reflect their listing strategies. \textbf{Myriad} exhibits high relative overlap despite originating few events, with most relations downstream, consistent with its editorial integration of prediction markets in news websites. At the opposite extreme, \textbf{Limitless} shows minimal overlap, reflecting a focus on short-horizon, option-style price predictions. \textbf{PredictIt} similarly exhibits limited overlap, driven by its emphasis on highly granular, U.S.\ state-level political markets that rarely receive parallel listings elsewhere.

\subsection{Temporal Evolution of Cross-Platform Semantic Overlap}

\begin{figure}[htbp]
    \centering
    \includegraphics[width=\linewidth]{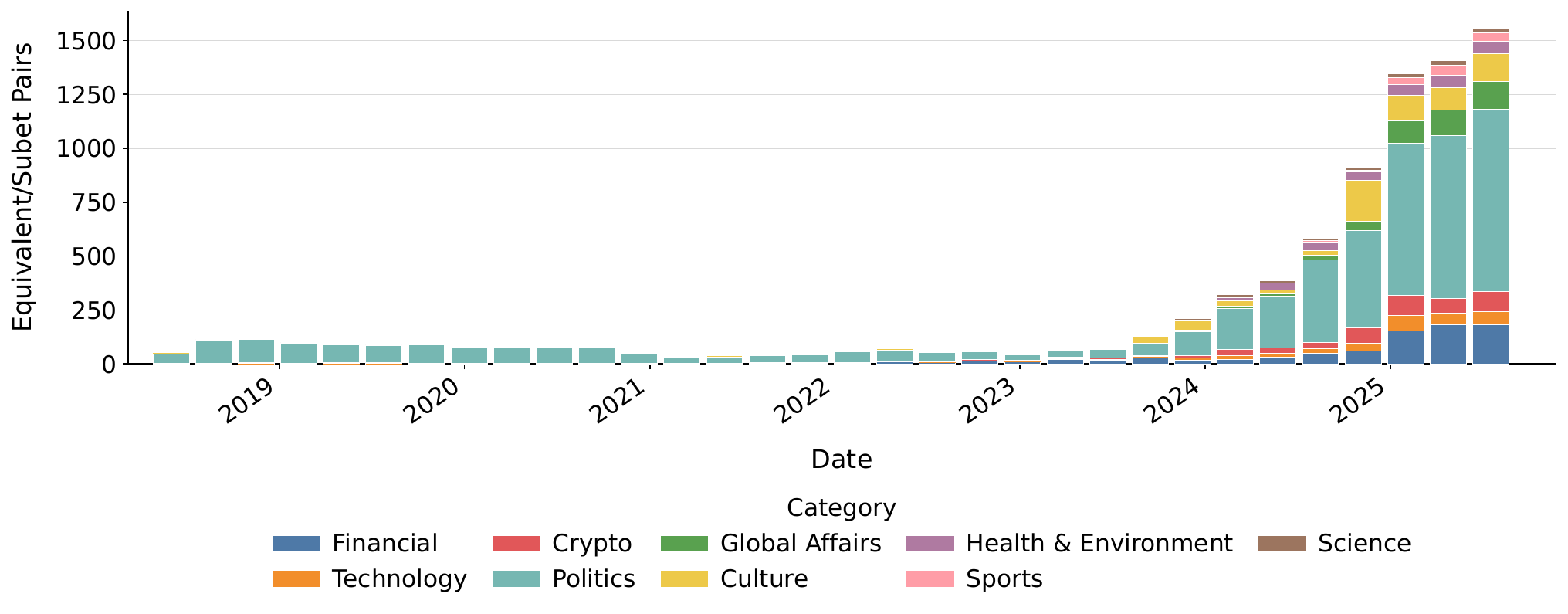}
    \caption{
        Number of equivalent or subset market pairs over time, grouped by category (top) and platform (bottom).
        The stacked bars indicate concurrent market activity in 42-day intervals.
        Equivalent pairs are counted once; multi-relations contribute one pair per direction.
    }
    \label{fig:histogram_activity}
\end{figure}

Figure~\ref{fig:histogram_activity} shows the temporal evolution of cross-platform semantic overlap, measured as the number of equivalent or subset market pairs active in fixed 42-day windows and grouped by category.

From 2018 through late 2022, overlap is negligible: matched pairs rarely exceed 100 per window and are almost entirely confined to political markets, reflecting an early-stage ecosystem with few concurrently active platforms and limited coverage outside major elections. Beginning in early 2024, cross-platform duplication increases sharply and expands beyond politics into finance, crypto, sports, culture, and global affairs. By 2025, windows routinely exceed 1{,}200–1{,}500 matched pairs, surpassing the cumulative overlap observed over the entire prior history. Although political events remain dominant, the rapid diversification across categories indicates that multi-platform duplication is no longer election-driven but has become systemic.

This acceleration coincides with Polymarket’s growth during the 2024 U.S.\ election cycle, the launch of new on-chain venues (e.g., Limitless, Truemarkets, Myriad), and regulatory changes that expanded political-market participation on Kalshi. As platforms increasingly list semantically identical markets independently, the ecosystem now generates large volumes of redundant liquidity.

In the absence of semantic interoperability, this redundancy manifests as fragmentation: equivalent markets trade as distinct assets across venues, leading to inconsistent pricing, delayed information propagation, and persistent arbitrage spreads, quantified in the next subsection.

\subsection{Price Divergence and Arbitrage Dynamics}

We quantify the economic consequences of cross-platform semantic overlap by measuring the prevalence, magnitude, and persistence of execution-aware deviations from no-arbitrage parity across platforms, and by characterizing the resulting arbitrage dynamics. Due to data availability constraints, this analysis covers \textbf{7 out of the 10} platforms in our dataset for which reliable price time series, fee schedules, market structures, and execution-cost parameters can be reconstructed; the full platform coverage and cost assumptions are documented in Appendix~\ref{app:execution_costs_section}.

\begin{figure}[htbp]
    \centering
    \includegraphics[width=\linewidth]{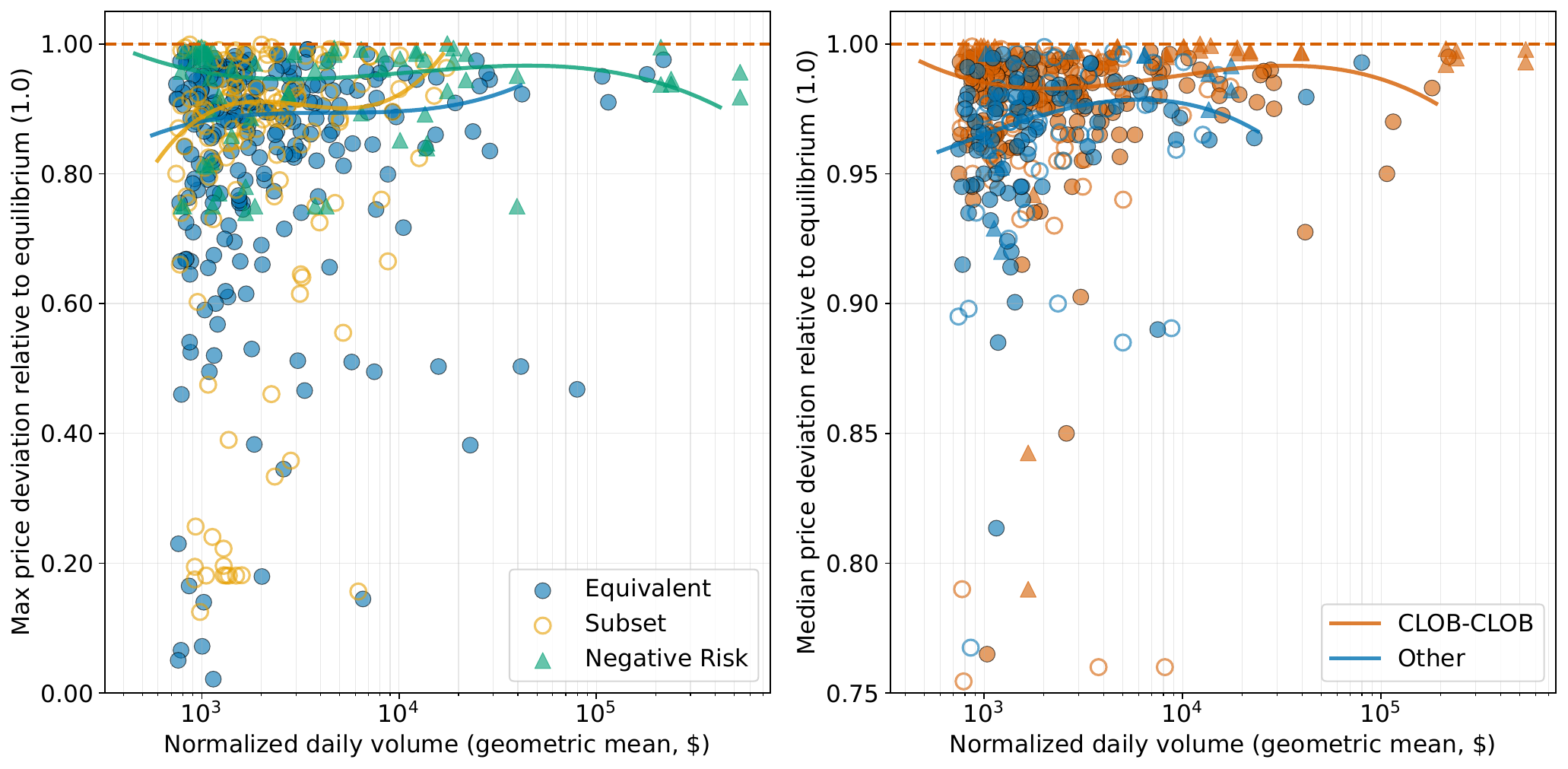}
    \caption{
        Price deviation from equilibrium as a function of effective liquidity for cross-platform arbitrage, shown for the top \textbf{1{,}000} relations by total trading volume.
        Markers denote arbitrage type (filled circle: equivalent; open circle: subset; triangle: negative risk).
        Colors indicate arbitrage type (left) or market structure (right).
        Dashed lines mark equilibrium, and solid curves show robust smoothed averages.
        Maximum price deviations are required to persist for at least one hour.
    }
    \label{fig:cross_platform_arbitrage_liquidity}
\end{figure}

\paragraph{Existence and Structure of Cross-Platform Price Divergence}

Figure~\ref{fig:cross_platform_arbitrage_liquidity} summarizes cross-platform price divergence along two complementary dimensions. The left panel reports the \emph{maximum} execution-aware deviation from parity for each arbitrage construction, requiring the deviation to persist for at least one hour, capturing the largest economically meaningful mispricing over the joint lifetime of the aligned markets. The right panel reports the \emph{median} deviation relative to equilibrium, reflecting the typical magnitude of price disagreement.

Both panels plot price deviations against effective liquidity, defined as the geometric mean of normalized daily trading volume across all participating binary markets. Maximum deviations remain large even at moderate liquidity levels and decline only gradually with volume, indicating that substantial mispricing occurs in actively traded markets. Median deviations are smaller but systematically positive: even among the most liquid constructions, prices typically remain 2--4\% away from execution-adjusted parity.

Taken together, the panels show that cross-platform price divergence is not driven solely by rare spikes. Instead, markets exhibit persistent directional disagreement: large deviations arise episodically (left), while smaller but economically meaningful gaps persist for a substantial fraction of time (right). Greater liquidity reduces but does not eliminate either effect.

\begin{figure}[htbp]
  \centering
  \includegraphics[width=\linewidth]{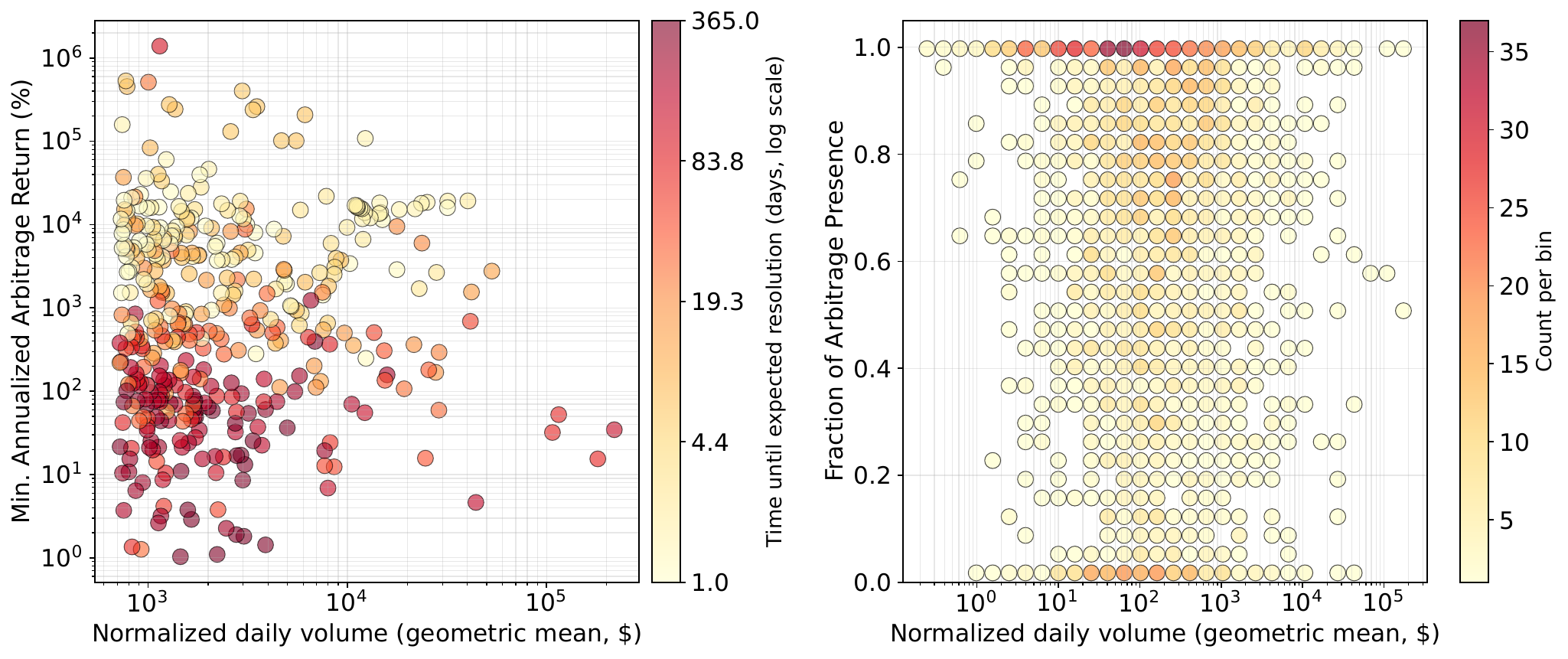}
    \caption{
        Relationship between market liquidity and conditional arbitrage characteristics for equivalent prediction market pairs.
        \textbf{Left:} Maximum guaranteed (worst-case) annualized arbitrage return versus effective liquidity (geometric mean of normalized daily volume), with color indicating the median time to resolution.
        \textbf{Right:} Fraction of time during which an execution-aware, risk-free arbitrage opportunity exists, shown as a function of liquidity using log-volume binning.
    }
  \label{fig:liquidity-arbitrage-apy-arbshare}
\end{figure}

\paragraph{Temporal Persistence and Directionality of Price Divergence}

Figure~\ref{fig:liquidity-arbitrage-apy-arbshare} characterizes the temporal structure of price divergence for \textbf{equivalent-market arbitrages}, incorporating platform-specific fees and typical bid--ask spreads for CLOB-based venues. The left panel plots the \emph{maximum guaranteed annualized arbitrage return}, computed under worst-case (latest) resolution assumptions, against effective liquidity. High-yield opportunities arise across the liquidity spectrum but are typically short-lived, as reflected in the strong negative relationship between annualized return and median time to resolution. Consequently, large APYs are driven primarily by short market horizonts until resolution rather than substantial mispricing.

The right panel reports the fraction of time during which an execution-aware, risk-free arbitrage exists. Two regimes are apparent. A substantial mass of market pairs exhibits values near one, indicating persistent price divergence over nearly the entire overlap period. By contrast, another group is concentrated near zero, reflecting cases in which execution costs—most notably platform fees on venues such as Futuur—fully absorb nominal price differences and eliminate arbitrage despite visible mid-quote deviations.

Persistent arbitrage is driven primarily by \emph{directional disagreement}, with one platform pricing the event consistently higher than the other for most of the market lifetime. A smaller subset reflects large-amplitude price fluctuations that remain above execution costs even as direction occasionally reverses, rendering arbitrage continuously capturable. Overall, the figure shows that cross-platform price divergence is predominantly structural rather than transient.

\paragraph{Magnitude of Execution-Aware Yield}

\begin{figure}[htbp]
    \centering
    \includegraphics[width=\linewidth]{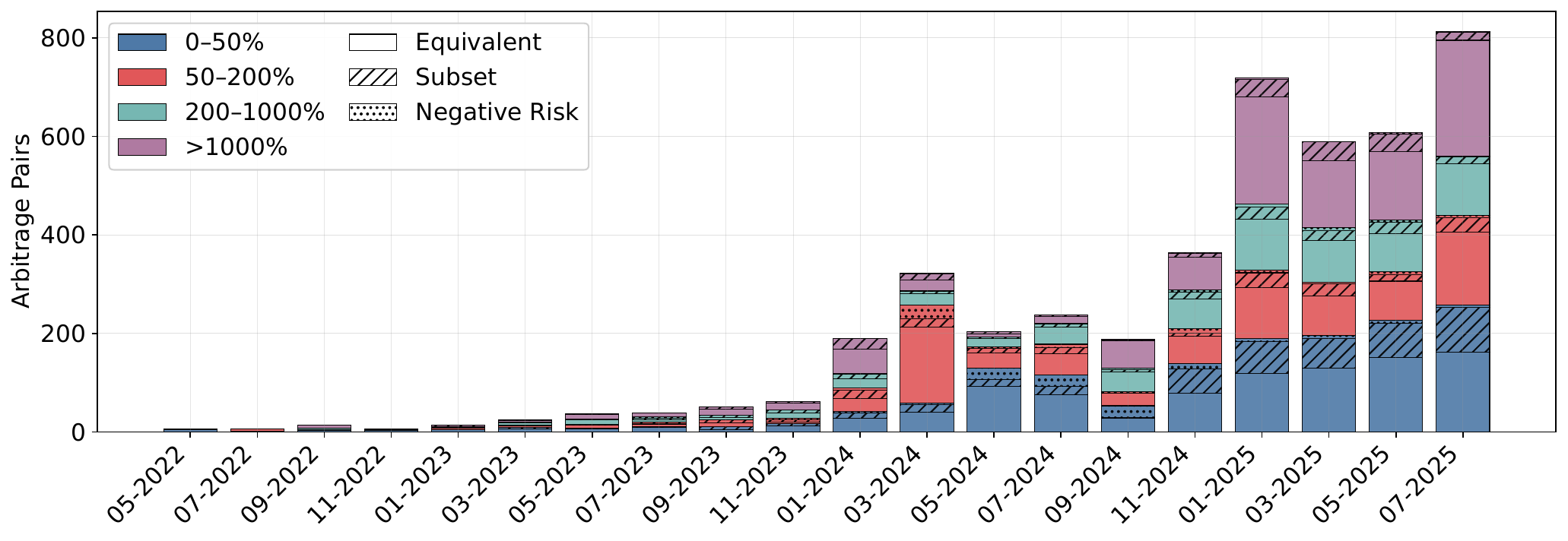}
    \caption{
        Distribution of arbitrage opportunities by their maximum execution-aware annualized return, requiring persistence of at least one hour.
    }
    \label{fig:apy-histogram-over-time}
\end{figure}

Figure~\ref{fig:apy-histogram-over-time} reports the distribution of arbitrage opportunities by their \emph{maximum execution-aware annualized return}, accounting for platform fees and typical bid--ask spreads (Appendix~\ref{app:execution_costs_section}). While many opportunities imply modest yields, a substantial mass exceeds 200\% APY, and a non-trivial fraction exceeds 1{,}000\%.

To contextualize these magnitudes, we simulate a naive, fully mechanical strategy starting in 2022. At each point in time, the trader enters the single highest-yield equivalent-market arbitrage available and holds the position until resolution, without switching or timing optimization. Even under this restrictive policy and after accounting for all execution frictions, the strategy yields a cumulative return of \textbf{1{,}218.66\%} over 800 days, across 15 completed trades.

\subsection{Case Study: The 2024 U.S.\ Presidential Election}
\label{sec.case_study}

The 2024 U.S.\ presidential election provides a canonical test for cross-platform price consistency and illustrates the mechanisms underlying the patterns documented above. It is the most liquid prediction-market event to date, listed across nearly all major centralized and decentralized venues, and outcome-relevant information becomes effectively common knowledge immediately following election-night network calls. Consistent with our earlier findings, substantial and persistent cross-platform price divergence is observed even in this information-rich environment. Such dispersion cannot plausibly be attributed to heterogeneous beliefs; instead, it isolates structural frictions that impede price convergence. Price trajectories for this event are reported in Appendix~\ref{app:us_election_comparison}.

A primary source of divergence arises from differences in resolution semantics for the Trump-YES payoff. Polymarket and Limitless resolve based on media network calls, whereas Kalshi and Myriad resolve based on the inauguration outcome. In terms of the atomic outcome sets defined in Section~\ref{sec.formal_market_model}, this induces a strict subset relation: every state in which Trump is inaugurated is contained in the set of states in which major networks call the race in his favor, but not vice versa.\footnote{The two sets do not coincide exactly: there exist conceivable sequences in which a candidate is declared the loser by major networks yet ultimately assumes office.} Other platforms (e.g., Futuur, Seer) exhibit limited election-night activity, reflecting partial liquidity withdrawal.

\begin{figure}[htbp]
    \centering
    \includegraphics[width=\linewidth]{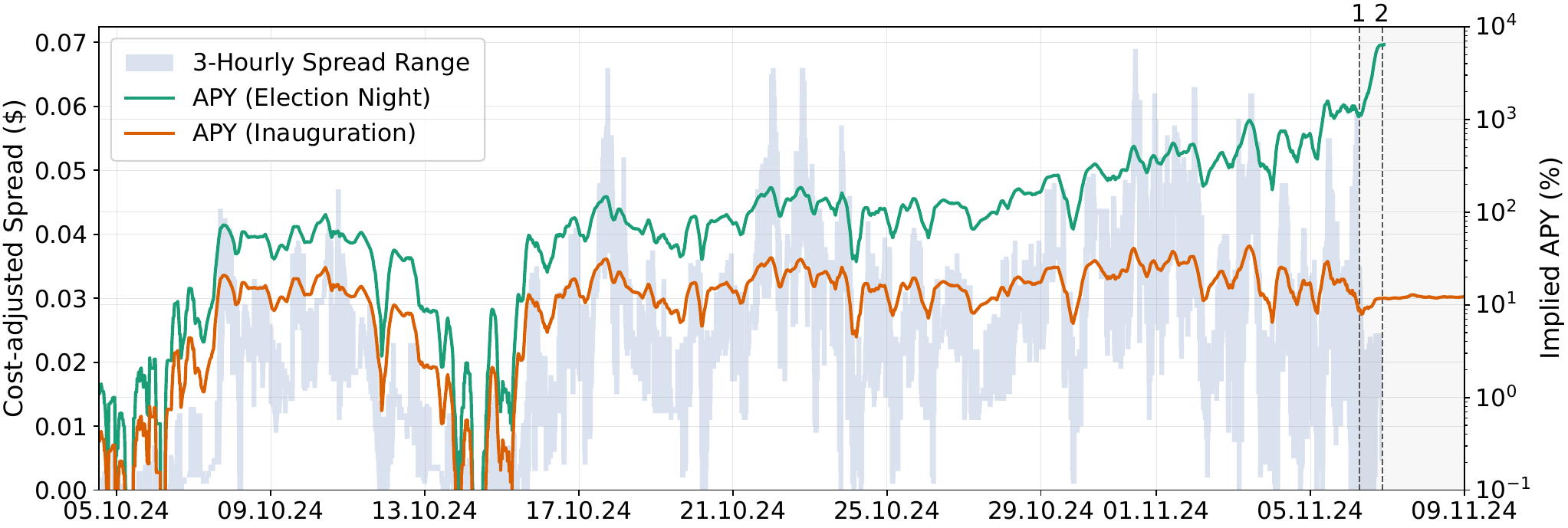}
    \caption{
        Execution-adjusted Kalshi--Polymarket price divergence for the 2024 U.S.\ presidential election.
        The left axis shows the rolling three-hour min--max spread.
        The right axis reports the implied annualized return under alternative settlement horizons, corresponding to resolution at election-night network calls (optimistic) and at inauguration (pessimistic).
        Vertical dashed lines mark (1) the first major election-night call and (2) Polymarket’s market-closure timestamp.
    }
    \label{fig:kalshi-polymarket-apy}
\end{figure}

\subsubsection{Polymarket–Kalshi Subset–Superset Arbitrage}

Among all platform pairs, Polymarket and Kalshi provide the most liquid overlapping instantiation of the 2024 U.S.\ presidential election (Polymarket: \$3.3B cumulative volume; Kalshi: \textgreater\$1B). During the observation period, Polymarket operated a zero-fee central limit order book with a \$0.001 tick size, while Kalshi employed \$0.01 ticks and charged no transaction fees for this market. All price comparisons therefore use the execution-adjusted price measure $P_Y(m)\pm\delta(m)$ defined in Section~\ref{sec.formal_market_model}, where $\delta(m)$ aggregates bid--ask spreads, platform fees, gas costs, and tick-size constraints.\footnote{Given the substantial depth observed on both venues, slippage is assumed negligible over the relevant horizons.}

Because Polymarket’s contract corresponds to a strict superset of Kalshi’s ($f_{\text{Kalshi}}\subset f_{\text{Poly}}$), the pair induces a valid subset--superset construction: taking YES on Polymarket and NO on Kalshi spans the full atomic outcome space. Figure~\ref{fig:kalshi-polymarket-apy} reports rolling three-hour min--max price differences, retaining only deviations persisting for at least thirty minutes. Prior to the election-night call, execution-adjusted spreads average approximately \$0.03 and reach up to \$0.07 for sustained intervals. Depending on the assumed capital lockup horizon release upon Polymarket’s call-based closure versus Kalshi’s inauguration-based resolution—these gaps correspond to deterministic annualized returns ranging from single-digit percentages to several hundred percent.

Regulatory segmentation contributes to the persistence of these deviations but does not fully account for them. Kalshi primarily serves U.S.\ participants, whereas Polymarket operates under different regulatory constraints, leading to partially segmented trader populations. Despite this segmentation, substantial overlap in information and market participation exists, yet cross-platform price convergence remains incomplete.

Even in this uniquely liquid and information-rich setting, cross-platform prices fail to converge. The Polymarket–Kalshi comparison illustrates that fragmentation in prediction markets is structural rather than incidental, arising from heterogeneous resolution semantics, institutional segmentation, and persistent limits to cross-platform arbitrage.

\section{Discussion}

Cross-platform price divergence in prediction markets persists even in highly liquid and information-rich settings. This does not contradict market efficiency once enforceability and capital-allocation constraints are made explicit. Classical no-arbitrage arguments presuppose fungible assets with enforceable identity and capital-efficient netting, neither of which holds for prediction-market claims instantiated independently across platforms. While convergence could, in principle, be induced by holding offsetting positions across venues until resolution, such positions are capital-intensive and ill-suited to short-horizon price alignment. Persistent divergence thus reflects constrained arbitrage capacity rooted in semantic non-fungibility rather than irrational pricing.

These findings imply that prediction markets aggregate information locally rather than globally. Prices are internally meaningful within platforms but are not directly comparable across venues in the absence of a shared, machine-verifiable notion of event identity. Fragmentation is most pronounced for long-lived, high-salience events that are concurrently listed across platforms—precisely where global information aggregation would be most valuable. As a result, a single, well-defined “market probability” for an event does not exist in the current ecosystem.

By contrast, markets for standardized tokens exhibit rapid cross-chain price alignment, highlighting that interoperability at the level of asset definition is a prerequisite for convergence. No analogous mechanism currently exists for prediction-market claims, whose payoff equivalence cannot be enforced or netted across venues. While our empirical analysis relies on approximate measures of trading volume and execution frictions, and abstracts from rare edge cases in resolution semantics, these choices if anything bias results toward understating fragmentation. The central conclusion is therefore robust: absent enforceable cross-platform event identity, price fragmentation in prediction markets is structural rather than transient.

\section{Related Work}

The theoretical foundation for prediction markets as mechanisms for aggregating dispersed information is well established. Market-scoring-rule designs, most notably the LMSR introduced by Hanson~\cite{repec:buc:jpredm:v:1:y:2007:i:1:p:3-15} and his earlier work on combinatorial markets~\cite{Hanson2003}, provide bounded-loss automated market makers whose prices admit a probability-like interpretation. Wolfers and Zitzewitz~\cite{Wolfers2004} formalize the economic rationale for treating contingent-claim prices as forecasts of future events and survey empirical evidence demonstrating predictive accuracy across political, economic, and corporate domains.

Empirical studies to date examine predominantly \emph{single-platform} behavior. Sirolly et~al.~\cite{Sirolly2025} analyze trading activity on Polymarket and propose a graph-based method for detecting wash trading. Saguillo et~al.~\cite{saguillo_et_al:LIPIcs.AFT.2025.27} document within-platform arbitrage opportunities in Polymarket markets. Ng et~al.~\cite{Ng2025} investigate price discovery and lead–lag dynamics across major platforms during the 2024 U.S.\ presidential election. While these studies yield platform-specific insights, they do not provide an ecosystem-level perspective: comparisons across venues are limited in scope, and semantic equivalence among markets is assumed rather than established.

Findings from adjacent domains highlight how execution frictions limit cross-market price alignment even when the traded assets are \emph{fully fungible}. Cryptocurrency markets exhibit persistent cross-exchange deviations due to segmentation and limits to arbitrage~\cite{repec:eee:jfinec:v:135:y:2020:i:2:p:293-319}. Cross-chain arbitrage studies similarly show that heterogeneous finality guarantees, bridging delays, and capital-mobility constraints sustain long-lived price gaps for identical tokens~\cite{oz2025crosschain}. Prediction markets inherit many of these frictions but introduce a domain-specific challenge: \emph{semantic non-fungibility}. Because platforms independently define event descriptions, oracle semantics, and cutoff rules, economically identical contingent claims lack a canonical identifier. Prior work does not model this semantic friction and therefore cannot explain observed cross-platform price divergence.

\section{Conclusion}

This paper shows that prediction markets violate the Law of One Price for structural reasons that go beyond transaction costs, latency, or liquidity. Economically equivalent contingent claims trade at persistent price differences even when information is common knowledge and markets are highly liquid. The underlying cause is semantic non-fungibility: prediction-market claims lack a shared, machine-verifiable notion of event identity. In the absence of such identity, liquidity fragments across venues and arbitrage cannot enforce price parity.

We provide the first ecosystem-scale empirical analysis of this phenomenon, spanning over 100{,}000 events across ten major platforms. By introducing a formal notion of cross-platform event identity and constructing a large, human-validated dataset of semantically aligned markets, we show that cross-platform price divergence is both economically significant and persistent. Fragmentation is concentrated in long-lived, high-salience events, implying that prices aggregate information locally within platforms but fail to do so globally across the ecosystem.

Unlike fungible assets, prediction-market claims cannot be proven equivalent, netted, or offset atomically across venues. As a result, arbitrage positions require capital to be committed until resolution, and price convergence would require a substantial volume of capital to be allocated to such long-lived cross-platform positions. In practice, this limits the ability of arbitrage to discipline prices even in liquid markets. Rebalancing alone cannot eliminate divergence: without enforceable outcome interoperability, price misalignment persists.

These limitations reflect intrinsic heterogeneity in event specification rather than methodological artifacts. More broadly, our findings indicate that prediction markets fail to converge not because of insufficient information or incentives, but because they lack a shared, machine-verifiable language for contingent claims. Establishing such semantic foundations is therefore a prerequisite for prediction markets to function as mechanisms for global information aggregation.

\begin{credits}
\subsubsection{\ackname} We thank Burak Öz for valuable feedback on earlier drafts of this manuscript.

\subsubsection{\discintname}
The authors declare no financial or non-financial competing interests and no funding from any of the platforms analyzed in this work.
\end{credits}

%
%

\bibliographystyle{splncs04}
\bibliography{mybibliography}

\newpage

\appendix

\section{Overview of Analyzed Prediction Markets}

\begin{table}[!ht]
\centering
\scriptsize
\renewcommand{\arraystretch}{1.0}
\setlength{\tabcolsep}{3pt}
\newcommand{\fixheight}{\rule{0pt}{2.8ex}} 

\caption{Overview of Major Prediction Market Platforms}
\label{tab:platforms_overview}
\begin{tabularx}{\linewidth}{@{}p{1.8cm}p{1.7cm}p{1.0cm}p{3.3cm}X@{}}
\toprule
\textbf{Platform} & \textbf{Chain} & \textbf{Launch} & \textbf{Pricing Mechanism} & \textbf{Resolution Mechanism} \\
\midrule

Polymarket \fixheight &
Polygon &
2020 &
Pre-2022 CPMM; post-2022 CLOB (no fees). &
UMA Optimistic Oracle, 2 h challenge window, dispute process. \\

Kalshi \fixheight &
Off-chain &
2021 &
CLOB. &
Outcome Review Committee. \\

PredictIt \fixheight &
Off-chain &
2014 &
CLOB. &
Operator-resolved. \\

Futuur \fixheight &
Off-chain &
2017 &
Hybrid CLOB / LMSR. &
Operator-resolved, partly automated. \\

Augur v1 \fixheight &
Ethereum &
2018, 2020 &
CLOB (0x Protocol). &
REP-based decentralized oracle with dispute. \\

Myriad \fixheight &
Abstract, Linea, Celo &
2025 &
CPMM. &
Operator-resolved. \\

Limitless \fixheight &
Base &
2024 &
Mixed (CLOB, CFMM). &
Automated oracle resolution via Pyth Network. \\

Truemarkets \fixheight &
Base &
2025 &
Aggregated liquidity (CFMM, Uniswap v3). &
Optimistic oracle resolution, token vote, attester jury. \\

Seer \fixheight &
Gnosis, Ethereum &
2024 &
Aggregated liquidity (Swapr v3). &
Reality.eth oracle, Kleros arbitration. \\

Omen/Presagio \fixheight &
Gnosis &
2020, 2025 &
CPMM. &
Reality.eth oracle, Kleros arbitration. \\

\bottomrule
\end{tabularx}
\end{table}

\newpage

\section{Market Categorization Prompt}
\label{app:categories}

\noindent\textbf{Description.}  
This prompt instructs the language model to assign each prediction-market event 
to one of twenty predefined semantic categories, returning only the numerical 
category identifier to ensure consistent downstream filtering and retrieval.

\begin{verbatim}
You are a prediction market category classifier. 
Given a prediction market description, assign it to the most
appropriate category from the numbered list provided.

Available categories (respond only with the number):

Financial:
1. Markets & Trading: Stocks, trading, market movements, IPOs
2. Economics & Policy: Economic indicators, central banks, fiscal
   policy

Technology:
3. Technology & Software: Hardware, software companies, products
4. Digital Platforms: Social media, online services, internet
   platforms
5. AI & Computing: AI, ML, algorithms, automation

Crypto:
6. Digital Assets: Cryptocurrencies, tokens, NFTs
7. Blockchain & Web3: Blockchain tech, DeFi, web3 infrastructure

Political:
8. Elections & Politics: Elections, campaigns, voting
9. Governance & Policy: Government actions, regulations,
   legislation

Global Affairs:
10. International Diplomacy: International relations, treaties,
    diplomacy
11. Conflicts & Security: Wars, military actions, security issues

Culture:
12. Media & Entertainment: Movies, music, celebrity, shows
13. Society & Culture: Social trends, cultural issues,
    demographics
14. Legal & Justice: Court cases, legal developments,
    justice system

Health & Environment:
15. Health & Medical: Healthcare, diseases, medicine, pharma
16. Environment & Climate: Climate change, pollution,
    sustainability

Other:
17. Sports: All sports, tournaments, athletics
18. Gaming & E-Sports: Video games, gaming industry,
    competitive gaming
19. Science & Research: Scientific discoveries, academic
    research
20. Space & Exploration: Space missions, astronomy, aerospace

Rules:
1. Respond ONLY with the category number (1–20)
2. Provide no explanation or additional text
3. If unsure, choose the most relevant category number
4. Be consistent in your categorization

Prediction market description:

"{description}"

Category number:
\end{verbatim}

\newpage

\section{Validation of Semantic Retrieval Recall}
\label{app:rank_distribution}

\begin{figure*}[htbp]
  \centering
  \includegraphics[width=\textwidth]{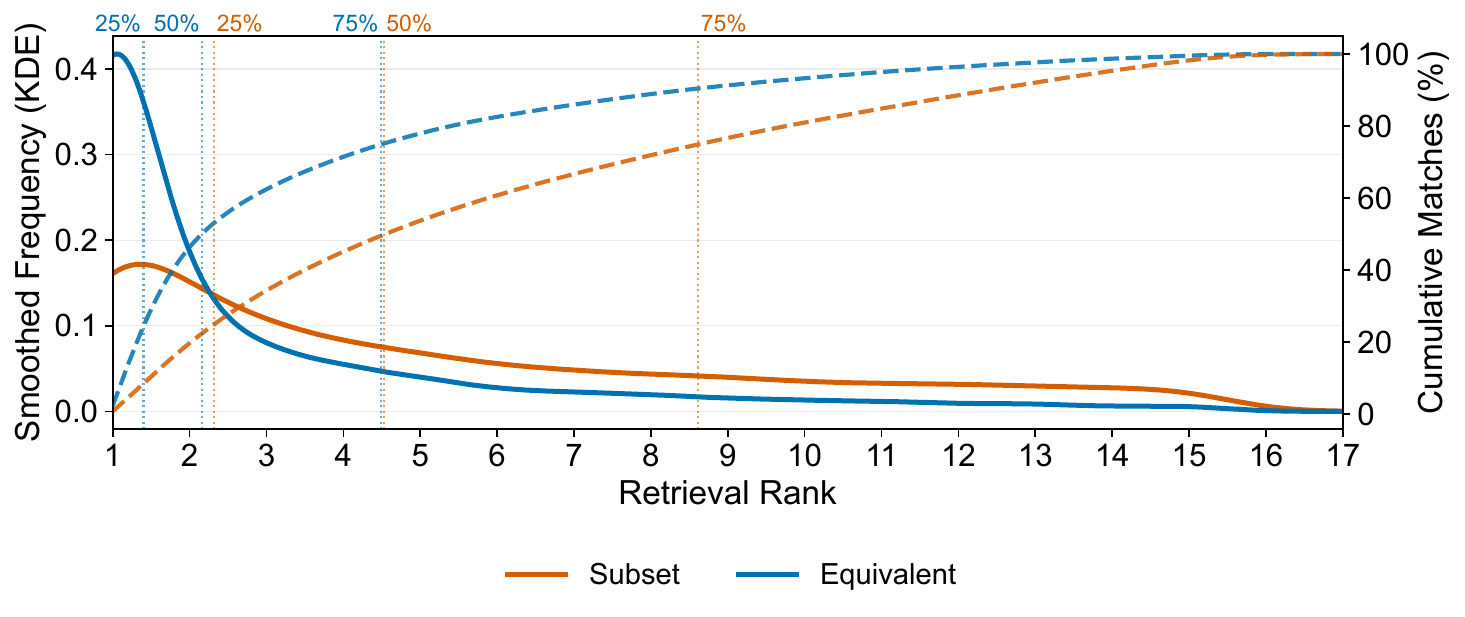}
  \caption{
    Empirical rank distribution of true matches across all platforms.  
    Below one in 1000 of equivalent and subset relations fall outside the top $k=16$ nearest neighbors during the semantic search, validating the retrieval depth.
  }
  \label{fig:rank-distributions}
\end{figure*}

\newpage

\section{LLM Prompt for High Recall Filtering}
\label{app:filtering_prompt}

\begin{verbatim}
You are an expert in prediction-market semantics.
Determine whether each candidate market refers to the same
real-world event as the target market — in other words,
whether they would resolve with the same outcome in all
relevant situations.

Definitions:
- Equivalent: both markets resolve identically in every
possible case.
- Subset: whenever the candidate resolves to YES, the target
would also resolve to YES, but not necessarily the reverse.
- Independent: any other relationship.

Output only candidates classified as Equivalent or Subset.

### Target Market
{target_document_text}

### Candidate Markets
Each entry includes: Rank, Description.

{similar_documents_text}

### Output Format
One line per match:
<rank>,<classification>

Example:
1,equivalent
4,subset
\end{verbatim}

\newpage

\section{LLM Prompt for Equivalence-Class Validation}
\label{app:equivalence_prompt}

\noindent\textbf{Description.}  
This prompt instructs the language model to determine which markets in a 
pre-selected candidate group are truly semantically equivalent, based on whether 
they would always resolve identically across all plausible real-world scenarios.

\medskip

\begin{verbatim}
You are checking whether a candidate prediction-market event is 
functionally equivalent to a reference event

DEFINITION OF EQUIVALENCE:
Two events are equivalent if they concern the same underlying
real-world phenomenon and would resolve the same way in all
realistic scenarios. Differences in wording, formatting,
outcome granularity, or detail level do not matter unless
they *change the resolution*.

BENEFIT OF THE DOUBT RULE:
If details are unspecified, assume standard conventions 
e.g., typical definitions, common measurement practices).  
If one event names a widely accepted data source and the
other does not, treat them as equivalent unless that
specification would realistically change the resolution.

EDGE-CASE HANDLING:
If one event explicitly describes how edge cases are resolved
and the other omits them, treat them as equivalent unless the
omission creates a realistic, impactful difference in final
resolution.

OUTCOME-SET FLEXIBILITY:
Outcome sets do NOT need to match.  
A finer-grained market is equivalent to a Yes/No market if one
of its outcomes corresponds exactly to the Yes-condition. Overlap
of a single outcome is sufficient. A specific-entity Yes/No market
is equivalent to a multi-option market if the entity appears as
one of the possible outcomes.

WHEN TO OUTPUT 1 (equivalent):
- Both events ask about the same real-world fact or phenomenon, AND
- There is **no explicit, impactful contradiction** in resolution
rules, AND
- There is no clear, realistic scenario where the resolutions
would differ.

WHEN TO OUTPUT 0 (not equivalent):
- There exists a concrete, plausible situation in which the two
events would certainly resolve differently.

IGNORE:
Contrived edge cases, stylistic differences, alternate but
standard data sources, or minor definitional noise.

OUTPUT FORMAT (strict):
Line 1: 1 or 0
Line 2: a 2–5 word reason.

REFERENCE EVENT:
{REFERENCE}

CANDIDATE EVENT:
{CANDIDATE}
\end{verbatim}

\newpage

\section{LLM Prompt for Subset-Relation Validation}
\label{app:subset_prompt}

\noindent\textbf{Description.}  
This prompt instructs the language model to evaluate whether each candidate 
market is a strict semantic subset of a given superevent, i.e., whether every 
scenario that satisfies the candidate market necessarily also satisfies the 
superevent.

\medskip

\begin{verbatim}
You are checking whether a candidate prediction-market event is a
valid SUBSET of a reference (superset) event.

DEFINITION OF SUBSET:
Event B is a subset of event A if, across all realistic real-world
scenarios, whenever event B would resolve as true (or to a specific
outcome), event A would also resolve as true (or to a corresponding
outcome). Event A may resolve as true in additional scenarios that
B does not cover. Small wording differences are acceptable; any
substantive mismatch that could cause B to resolve true while A
resolves false means B is NOT a valid subset of A.

BENEFIT OF THE DOUBT RULE:
If details are unspecified, assume standard conventions (e.g.,
typical definitions, common measurement practices). If one event
names a widely accepted data source and the other does not, treat
them as compatible unless that specification would realistically
change the resolution relationship.

EDGE-CASE HANDLING:
If one event explicitly describes how edge cases are resolved and
the other omits them, treat them as compatible unless the omission
creates a realistic, impactful difference that would break the subset
relationship.

OUTCOME-SET FLEXIBILITY:
Outcome sets do NOT need to match exactly. A finer-grained market can
be a subset of a Yes/No market if all outcomes that would make the
subset "true" correspond to the Yes-condition of the superset. A
specific-entity Yes/No market can be a subset of a multi-option market
if the entity appears as one of the possible outcomes and the subset's
Yes maps to that outcome.

WHEN TO OUTPUT 1 (valid subset):
- The candidate event asks about a more specific or narrower version
of the reference event's phenomenon, AND
- There is **no explicit, impactful contradiction** in resolution
rules, AND
- There is no clear, realistic scenario where the candidate would
resolve true while the reference would resolve false (or to a
different outcome).

WHEN TO OUTPUT 0 (not a subset):
- There exists a concrete, plausible situation in which the
candidate would resolve true while the reference would resolve
false (or to a different outcome), OR
- The candidate is actually broader than or independent of
the reference.

IGNORE:
Contrived edge cases, stylistic differences, alternate but standard
data sources, or minor definitional noise that doesn't affect
the subset relationship.

OUTPUT FORMAT (strict):
Line 1: 1 or 0
Line 2: a 2–5 word reason.

SUPERSET EVENT (Reference):
{REFERENCE}

SUBSET CANDIDATE:
{CANDIDATE}
\end{verbatim}

\newpage

\begin{landscape}
\section{Data Acquisition Details}

\centering
\footnotesize
\label{tab:data-acquisition}
\setlength{\tabcolsep}{4pt}
\begin{tabular}{@{}p{2.0cm}p{1.6cm}p{1.5cm}p{4.3cm}p{8.5cm}@{}}
\toprule
\textbf{Platform} & \textbf{Chain} & \textbf{\#Events} & \textbf{Data Source} & \textbf{Details} \\ 
\midrule
\textbf{Polymarket} & Polygon & 29'651 & Polymarket Gamma API; the-graph subgraph\footnotemark[1]; Polygonscan & For CPMM markets on-chain parsing required, respective timestamps from Polygon block time. \\[4pt]
\textbf{Augur v1} & Ethereum & 2'569 & Etherscan RPC & Start: contract creation; end: \texttt{MarketFinalized} event; timestamps via \texttt{proxy/eth\_getBlockByNumber}. \\[4pt]
\textbf{Limitless} & Base & 5'560 & Basescan; Limitless API & Start: contract creation (block time); markets with \texttt{volume = 0} excluded; API lists only active markets; CLOB markets therefore ignored as they lacked on-chain tracability. \\[4pt]
\textbf{Omen} & Gnosis & 12'846 & the-graph subgraph\footnotemark[2] & \texttt{creationTimestamp} from FPMM creation. Markets open < 30 days, otherwise discarded; many stale markets. \\[4pt]
\textbf{Myriad} & Abstract & 3'198 & Polkamarkets API &
Networks \texttt{network\_id} $\in \{274133, 2741, 274132\}$.
start: texttt{created\_at};
end: \texttt{expires\_at}. \\[4pt]
\textbf{Seer} & Gnosis & 878 & the-graph subgraph\footnotemark[3]; Gnosisscan & Closure via \texttt{payoutReported = true}; end timestamp from finalization transaction block time. \\[4pt]
\textbf{Truemarkets} & Base & 562 & Truemarkets API; Basescan & Start: contract creation block time; end; \texttt{statusUpdatedAt} (\texttt{status = 7}). \\[4pt]
\textbf{Kalshi} & Off-chain & 59'176 & Kalshi Trade API v2 & - \\[4pt]
\textbf{Futuur} & Off-chain & 43'910 & Futuur API &  - \\[4pt]
\textbf{PredictIt} & Off-chain & 4'381 & PredictIt API & Price data on request only. \\ 
\bottomrule
\end{tabular}

\footnotetext[1]{Subgraph ID: \texttt{81Dm16JjuFSrqz813HysXoUPvzTwE7fsfPk2RTf66nyC}.}
\footnotetext[2]{Subgraph ID: \texttt{9fUVQpFwzpdWS9bq5WkAnmKbNNcoBwatMR4yZq81pbbz}.}
\footnotetext[3]{Subgraph ID: \texttt{BMQD869m8LnGJJfqMRjcQ16RTyUw6EUx5jkh3qWhSn3M}.}

\vspace{0.5em}

\end{landscape}

\newpage

\section{Cross-Platform Price Series for the 2024 U.S. Election}
\label{app:us_election_comparison}

\begin{figure}[ht]
    \centering
    \includegraphics[width=\textwidth]{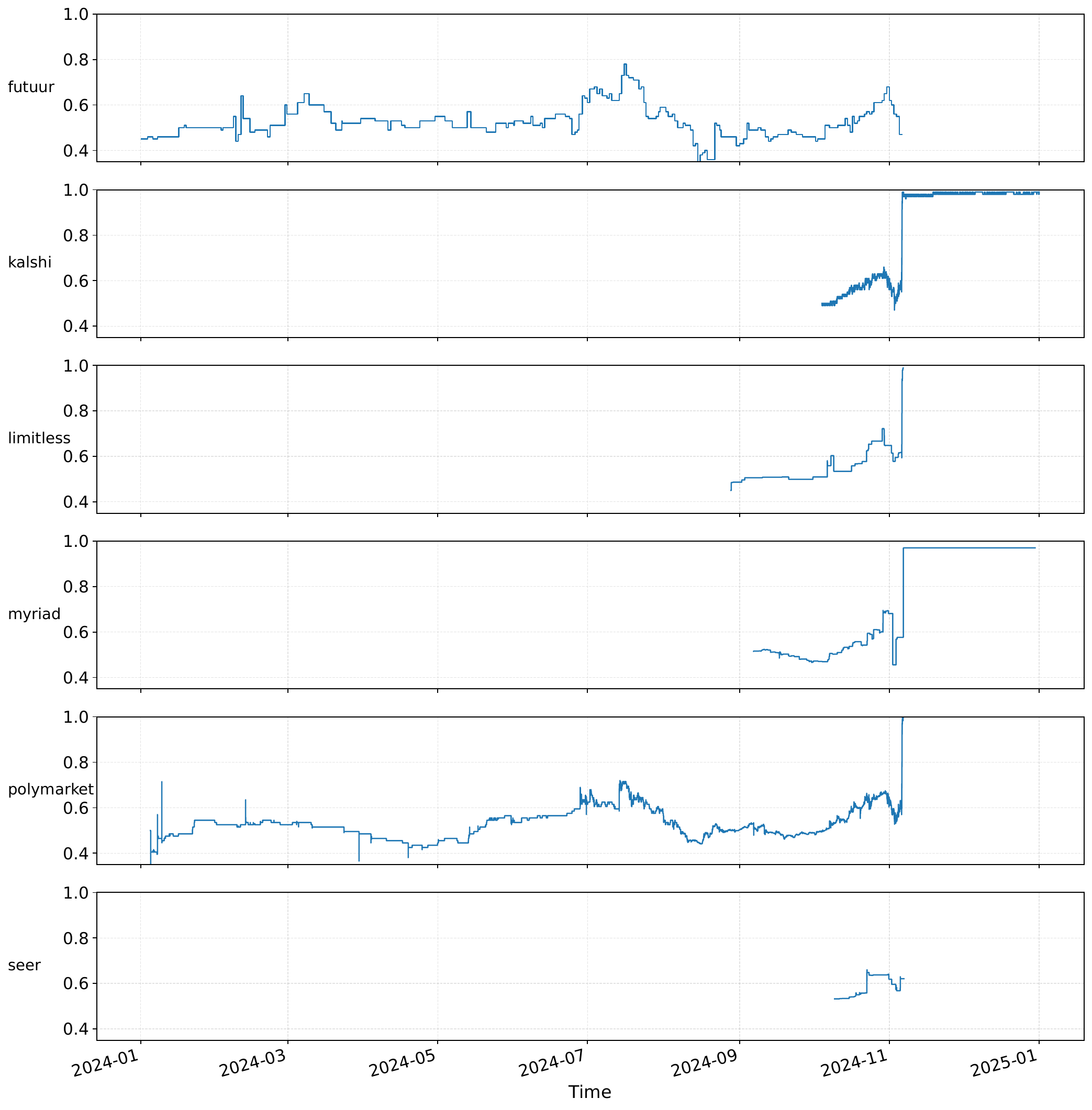}

    \caption{
        Time series of Trump–YES prices for the 2024 U.S.\ presidential election across major prediction-market platforms.
    }
    \label{fig:small-multiples-election}
\end{figure}

\newpage
\begin{landscape}

\section{Execution-Cost Assumptions and Platform Coverage}
\label{app:execution_costs_section}

\begin{table}[htbp]
\centering
\caption{Platforms included in the execution-aware price-divergence analysis and assumed execution-cost parameters. Reported fees and spreads reflect conservative, standardized assumptions used throughout the analysis to ensure cross-platform comparability.}
\label{tab:execution_costs}
\renewcommand{\arraystretch}{1.15}

\begin{tabular}{lcccc}
\toprule
\textbf{Platform} & \textbf{Market Structure} & \textbf{Fee Assumption} & \textbf{Spread Assumption} & \textbf{Notes} \\
\midrule
Kalshi      & CLOB & 1.5\% & \$0.01 & Dynamic fee schedule; conservatively fixed at 1.5\% \\
Polymarket  & CLOB & 0\%   & \$0.01 (0.01--0.99) & \$0.001 spread assumed near bounds \\
Polymarket  & CPMM & 2\%   & 0 &   \\
Futuur      & CLOB + LMSR & 6\% & \$0.01 &  \\
Omen        & CPMM & 2\%   & 0 &   \\
Myriad      & CPMM & 0--2\% & 0 & Market-specific fees considered \\
Truemarkets & AMM (Uni v3) & 0.8\% & 0 &  \\
Limitless   & CPMM & 1.5\% & 0 & Fees range 0.03--3\%; conservatively at 1.5\% fixed \\
\bottomrule
\end{tabular}
\end{table}

\end{landscape}

\end{document}